%% file: konstantopoulos_shucs_I.tex
\newcommand\Lsun{\hbox{L$_\odot$}}
\newcommand\Msun{\hbox{M$_\odot$}}
\newcommand\Zsun{\hbox{Z$_\odot$}}
\newcommand\kms{\hbox{km$\,$s$^{-1}$}}

\newcommand\one{\,{\sc i}}
\newcommand\two{\,{\sc ii}}

\newcommand\tmult{\multicolumn{2}{c}}
\newcommand\hst{\textit{HST}}
\newcommand\gal{\textit{GALEX}}

\newcommand\ie{\textit{i.\,e.}}
\newcommand\eg{e.\,g.}
\newcommand\cf{\textit{cf.}}

\newcommand\uub{$U_{336}$}
\newcommand\bb{$B_{450}$}
\newcommand\vb{$V_{606}$}
\newcommand\ib{$I_{814}$}
\newcommand\bvi{\textit{BVI}}
\newcommand\dndt{$\textup{d}N/\textup{d}t$}
\newcommand\ubvi{\textit{UBVI}}
\newcommand\fubvi{$f_\textit{\scriptsize UBVI}$}

\newcommand\mhimath{M_\textup{\scriptsize H\one}}

\newcommand\ha{H$\alpha$}

\newcommand\fir{\textup{\scriptsize FIR}}

\newcommand\reffmath{R_\textup{\scriptsize eff}}
\newcommand\n{NGC~}
\newcommand\ned{\textit{NED}}
\newcommand{\asec}{$^{\prime\prime}$}
\newcommand{\farcs}{\mbox{\ensuremath{.\!\!^{\prime\prime}}}}

\newcommand{\ygg}{\textit{Yggdrasil}}
\newcommand{\rgc}{$r_\textup{\scriptsize gc}$}

\documentclass[preprint2]{aastex}

\usepackage{graphicx,epstopdf,rotating,enumitem}
\usepackage[symbol*]{footmisc}

\shorttitle{SHUCS I:~Survey Description; NGC~4041}
\shortauthors{Konstantopoulos et al.}

\begin{document}

\title{The Snapshot \textit{Hubble} U-Band Cluster Survey (SHUCS). I. Survey Description and First Application to the Mixed Star Cluster Population of \n4041\footnote{
Based on observations made with the NASA/ESA Hubble Space Telescope, obtained at the Space Telescope Science Institute, which is operated by the Association of Universities for Research in Astronomy, Inc., under NASA contract NAS 5-26555. These observations are associated with program \#SNAP 12229.}.
}
\author{I.~S.~Konstantopoulos\altaffilmark{1,2},
L.~J.~Smith\altaffilmark{3},
A.~Adamo\altaffilmark{4},
E.~Silva-Villa\altaffilmark{5},
J.~S.~Gallagher\altaffilmark{6},
N.~Bastian\altaffilmark{7,8},
J.~E.~Ryon\altaffilmark{6},
M.~S.~Westmoquette\altaffilmark{9},
E.~Zackrisson\altaffilmark{10},
S.~S.~Larsen\altaffilmark{11},
D.~R.~Weisz\altaffilmark{12},
J.~C.~Charlton\altaffilmark{2}
}

\altaffiltext{1}{Australian Astronomical Observatory, PO Box 915, North Ryde NSW 1670, Australia; iraklis@aao.gov.au.}
\altaffiltext{2}{Department of Astronomy \& Astrophysics, The Pennsylvania State University, University Park, PA 16802, USA.}
\altaffiltext{3}{Space Telescope Science Institute and European Space Agency, 3700 San Martin Drive, Baltimore, MD 21218, USA.}
\altaffiltext{4}{Max-Planck-Institut for  Astronomy, K\"onigstuhl 17, D-69117 Heidelberg, Germany.}
\altaffiltext{5}{D\'epartement de Physique, de G\'enie Physique et d'Optique, and Centre de Recherche en Astrophysique du Qu\'ebec (CRAQ), Universit\'e Laval, Qu\'ebec, Canada.}
\altaffiltext{6}{Department of Astronomy, University of Wisconsin-Madison, 5534 Sterling, 475 North Charter Street, Madison WI 53706, USA.}
\altaffiltext{7}{Excellence Cluster Universe, Boltzmann-Strasse 2, 85748 Garching bei M\"unchen, Germany.}
\altaffiltext{8}{Astrophysics Research Institute, Liverpool John Moores University, Egerton Wharf, Birkenhead, CH41 1LD, UK}
\altaffiltext{9}{European Southern Observatory, Karl-Schwarzschild-Strasse 2, 85748 Garching bei M\"unchen, Germany.}
\altaffiltext{10}{Department of Astronomy, Stockholm University, Oscar Klein Centre, AlbaNova, Stockholm SE-106 91, Sweden.}
\altaffiltext{11}{Department of Astrophysics/IMAPP, Radboud University Nijmegen, P.O. Box 9010, 6500 GL Nijmegen, The Netherlands.}
\altaffiltext{12}{Department of Astronomy, Box 351580, University of Washington, Seattle, WA 98195, USA.}

\begin{abstract}

	We present the Snapshot \textit{Hubble} U-band Cluster Survey (SHUCS), a project aimed at characterizing the star cluster populations of ten nearby galaxies ($d<23~$Mpc, half within $\approx12$~Mpc) through new F336W (U~band equivalent) imaging from Wide Field Camera 3, and archival \bvi-equivalent data with the \textit{Hubble Space Telescope}. Completing the \ubvi\ baseline reduces the age-extinction degeneracy of optical colours, thus enabling the measurement of reliable ages and masses for the thousands of clusters covered by our survey. The sample consists chiefly of face-on spiral galaxies at low inclination, in various degrees of isolation (isolated, in group, merging), and includes two AGN hosts. This first paper outlines the survey itself, the observational datasets, the analysis methods,
and presents a proof-of-concept study of the large-scale properties and star cluster population of \n4041, a massive SAbc galaxy at a distance of $\approx23~$Mpc, and part of a small grouping of six giant members.  We resolve two structural components with distinct
stellar populations, a morphology more akin to merging and interacting systems. We also find strong evidence of a truncated, Schechter-type mass function, and a similarly segmented luminosity function. These results indicate that binning must erase much of the substructure present in the mass and luminosity functions, and might account for the conflicting reports on the intrinsic shape of these functions in the literature. We also note a tidal feature in the outskirts of the galaxy in \gal\ UV imaging, and follow it up with a comprehensive multi-wavelength study of \n4041 and its parent group. We deduce a minor merger as a likely cause of its segmented structure and the observed pattern of a radially decreasing star formation rate. We propose that combining the study of star cluster populations with broad-band metrics is not only advantageous, but often easily achievable thorough archival datasets. 
\end{abstract}


% =============================================
% 1. INTRODUCTION 
% =============================================
\section{Introduction}\label{sec:intro}
The launch of the \textit{Hubble Space Telescope} (\hst) over two decades ago started a revolution in the study of extragalactic star clusters. The discovery of large numbers of young compact clusters in star-forming galaxies led to the suggestion that they could be the present-day analogues of globular clusters \citep[see reviews of][]{whitmore03, larsen04b}. The question of whether these young clusters can survive for a Hubble time has still not been settled but their longevity appears to be critically dependent on environmental conditions within their host galaxies \citep{RdGGV,bastian11}. The installation of  Wide Field Camera 3 (WFC3) on \hst\ has vastly upgraded the imaging capabilities of the telescope shortward of $4000~$\AA. It is now much easier to measure the age and mass distributions of large populations of star clusters in galaxies, and address fundamental questions such as their long term survival chances. To obtain ages and extinctions for clusters younger than $\sim 2$~Gyr, it is essential to obtain photometry across the \ubvi\ baseline \citep{anders04}.  Prior to WFC3, \hst\ U~band imaging of sufficient depth and spatial coverage was feasible for only a few regions of nearby galaxies \citep[\eg][]{smith07regb, anders04b}, and only distant systems in their entirety \citep[\eg][]{oestlin03,adamo10a}.  

All local, late-type giant galaxies host populations of young and intermediate-age star clusters, often with masses and densities that rival globular clusters. It has been proposed that the vast majority of stars are formed in clusters but that most of these clusters ($\sim 90\%$) rapidly dissolve (Lada \& Lada 2003) due to the expulsion of residual gas from star formation. Clusters that survive this period of ``infant mortality'' can disrupt through stellar evolution,  two-body relaxation and the tidal field of the host galaxy on $\sim$~Gyr timescales \citep{bastian-gieles08}. There has been much debate in the literature over whether infant mortality exists, whether the later phases are mass-dependent or not \citep[\eg][]{lamers05, fall05, whitmore07, chandar10, bastian11}, and even if most stars do indeed form in clusters \citep{bressert10}. By conducting a survey of local star-forming galaxies, it will be possible to obtain large samples of clusters covering a wide range of masses and ages, and thus help answer several open questions. 

% Figure: n4041 HST
% ------------------------
\begin{figure*}[!t]
	\begin{center}

		\includegraphics[width=\textwidth,angle=0]{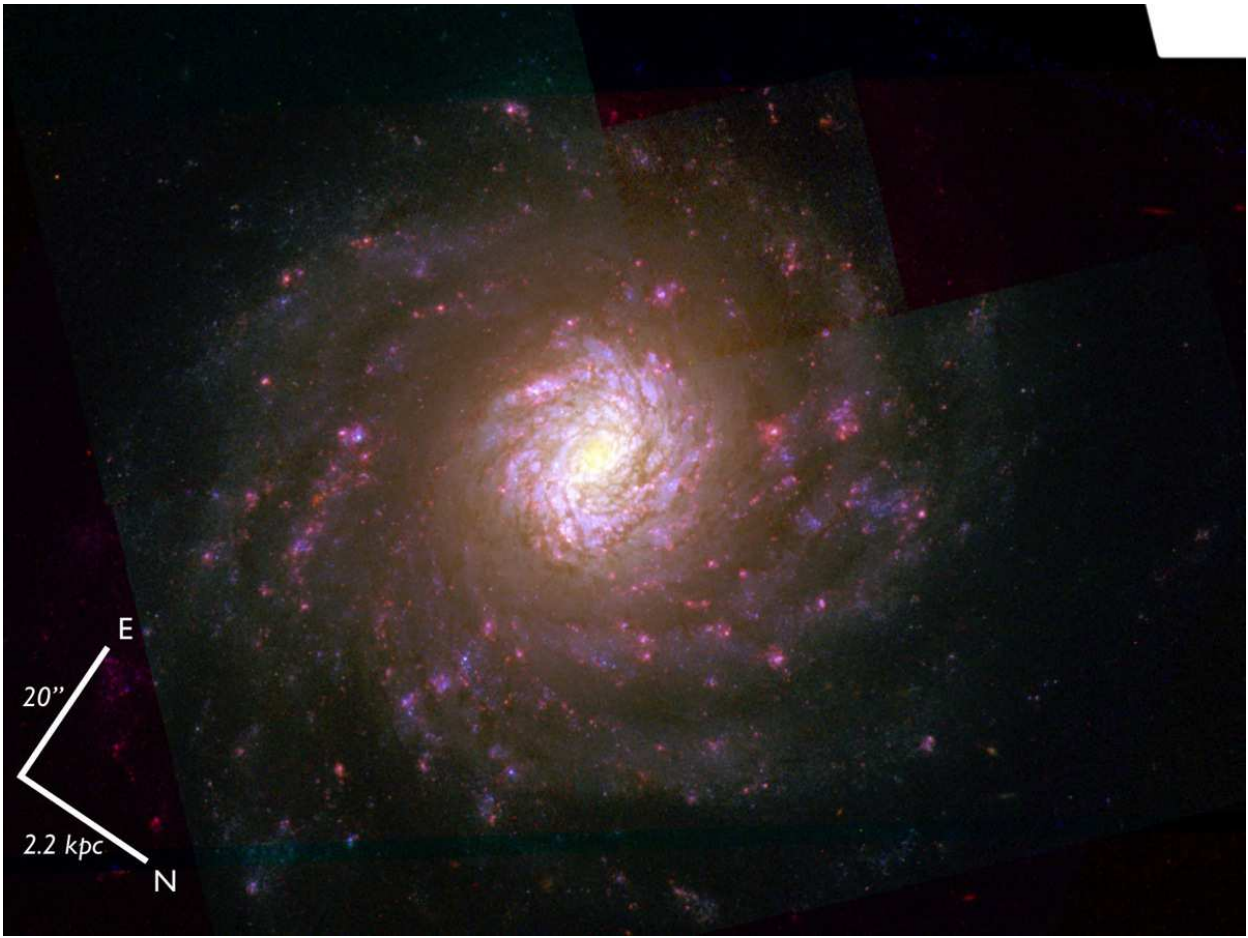}
		\caption{
			Colour composite imaging of \n4041, using all available 
			\hst\ imaging (\ubvi\ and H$\alpha$). The imprints of 
			various detectors are visible here, arising from uneven 
			spatial coverage across the optical baseline. The new 
			F336W imaging envelops the archival WFPC2 pointings, 
			while the ACS image leaves a trace of its chip gap. 
			The face of \n4041 exhibits a composite structure. 
			Pink H$\alpha$ bubbles trace the distribution of young 
			clusters along spiral arms that wind tighter in the inner 
			regions than the outer galaxy. A bright central component 
			seems to define this two-step structure, while the central
			peak is offset from the geometrical center, the center 
			of the outermost isophote, by $\approx1^{\prime\prime}$. 
		}\label{fig:hst}

	\end{center}
\end{figure*}
% ------------------------

In addition, the local environment has recently been suggested as a major contributor to cluster disruption \citep[\eg][]{elmegreen10,kruijssen12c}. The relationship between the formation of clusters and the properties of the host galaxy is, in fact, far from clear on both global and local scales, and a survey would help to investigate these relationships. So far, detailed studies of large numbers of clusters in small samples of galaxies \citep[\eg][]{meurer95,goddard10,adamo11,silvavilla11} or small numbers of clusters in large samples of galaxies \citep{larsen04,bastian08sfr,mullan11} have suggested that environments with a higher star-forming density form a higher fraction of their stars in clusters. 
This connection is also relevant to the link between galaxy interactions and increased star formation. Since star clusters can trace bursts of star formation, they can then provide a viable chronometer for the interaction history of a galaxy. 
 
Naturally, understanding the physics that govern these clusters is essential to utilizing them as tracers of star formation. After two decades of \hst-driven research, several cluster parameters are considered as standard, such as their distributions of luminosity and size. Perhaps most notably, many studies have investigated the star cluster mass distribution, with a common finding that it can be represented by a power law of index near $-2$ \citep[\eg][]{zhang99,RdG03,bik03}. More recent studies have, however, suggested that the true form is that of a Schechter function, a power-law distribution with an exponential truncation at high masses. {This reflects the mass function of Giant Molecular Clouds \citep[\eg][]{solomon87}, out of which star clusters form. The slope also appears to be a function of brightness, that is to say, the brighter the subsample, the steeper the extracted slope \citep{gieles10}}. In addition, preliminary indications suggest that the characteristic `Schechter mass', $M_*$, where the truncation occurs, depends on environment \citep{larsen09,gieles09a,bastian12}. Again, a large survey of clusters in a variety of host galaxies will be able to settle this issue.

Another parameter of interest is the star cluster size distribution and whether this is related to environment and/or age. Observations indicate that effective radii are constrained to a range of 0.5 -- 10 pc despite the large dynamic range in cluster mass \citep{pz10}. The observed radius distribution is well described by a log-normal distribution with a peak at 3--4~pc for both young clusters \citep{barmby06,bastian11} and old globular clusters \citep[\eg][]{jordan05}, with the exception of the ``faint-fuzzy'' clusters discovered by \citet{larsen00}. There are indications that the cluster core radius increases with age \citep[\eg][]{mackey03smc,mackey03lmc,remco07,bastian08cores,hurley10}. A survey of cluster sizes in different environments within galaxies will permit studies of the cluster size distribution as a function of age and environment.

From the above, it is clear that an extensive survey of a large sample of star clusters in a variety of environments, aimed at measuring their age, mass, and size distributions, will address many fundamental questions regarding their properties, survival rates, formation histories, and related environmental dependencies. In this paper, we describe such an endeavor: the Snapshot Hubble U-band cluster survey (SHUCS) combines new WFC3 F336W (U~band-equivalent) imaging with archival \hst~\bvi-equivalent imaging, to measure the properties of large samples of young clusters in nearby (mostly $d\lesssim12~$Mpc) star-forming galaxies. We present a full description of the survey in terms of sample definition and data reduction (Section~\ref{sec:survey}); the detection and photometry of star clusters, and the derivation of age, mass, and extinction (Section~\ref{sec:pipeline}). In the second part of this paper, we present a proof-of-concept study for \n4041, a bright \citep[$g=12.1~$mag,][]{rc3}  face-on SAbc galaxy with a double component disk (see Figure~\ref{fig:hst}). In Sections~\ref{sec:n4041}  and \ref{sec:clusters}, we discuss the large scale properties and the cluster populations of \n4041. We summarize this work, as well as our expectations for the full survey in Section~\ref{sec:summary}.

% =============================================
% 2. SURVEY DESCRIPTION
% =============================================
\section{The Survey: Target Selection and Data Reduction}\label{sec:survey}
\noindent To define our galaxy sample for WFC3-F336W imaging, we selected galaxies within 25~Mpc that have \bvi\ images available in the \hst\ archive. We restrict the sample to any galaxy imaged with the Advanced~Camera~for~Surveys~(ACS) or the Wide~Field~and~Planetary~Camera~2~(WFPC2), but imposed no constraints on image depth. We chose this distance limit and instrument set to ensure that individual star clusters are resolved and that deep WFC3~F336W imaging can be obtained with a 30 min exposure time. The depth is quantified as reaching $m_\textup{\scriptsize F336W}\approx26~$mag at acceptable error levels, as will be demonstrated in Section~\ref{sec:completeness}. We excluded dwarf irregular galaxies as they typically host very few clusters \citep[\eg][]{seth04}, but do not impose a strict lower mass limit on the dataset.

%% TABLE: SHUCS SAMPLE -----------------------------------------------
\begin{deluxetable}{lccl r@{.}l r@{.}l cl} 
\tabletypesize{\scriptsize}
%\rotate
\tablewidth{0pt} 
\tablecolumns{10}
\tablecaption{WFC3/UVIS Target List.\label{tab:targ}} 
\tablehead{
	\colhead{Name} & 				% C1
	\colhead{RA} & 					% C2
	\colhead{DEC} &					% C3
	\colhead{Morphology} &			% C4
	\tmult{Distance} &				% C5, 6
	\tmult{$\log(M_*)$} & 			% C7, 8
	\colhead{Ref.} & 				% C9
	\colhead{Other \hst\ data}\\	% C10
	& 										% C1
	{(h~m~s)} & 							% C2
	{($\circ$~$\prime$~$\prime\prime$)} &	% C3
	&										% C4
	\tmult{(Mpc)} &							% C5, 6
	\tmult{(\Msun)} &						% C7, 8
	&										% C9..10
}
\startdata 
\n247  & 00~47~10.49 & $-$20~46~09.00 & SAB(s)d     &~~~~3&6 &~~9&06 & [1] & F110W, F160W\,(N)\\
\n672  & 01~47~54.06 & $+$27~25~55.80 & SB(s)cd     &  8&1   &  9&43 & [2] & F658N\,(A)\\
\n891  & 02~22~32.90 & $+$42~20~45.80 & SA(s)b?     & 10&2   & 10&84 & [1] & F250W\,(A), F656N\,(W2), F160W\,(N)\\
\n925  & 02~27~05.14 & $+$33~34~54.50 & SAB(s)d     &  9&3   &  9&65 & [3] & F160W\,(N)\\
\n1003 & 02~39~16.40 & $+$40~52~20.40 & SA(s)cd     & 11&1   &  9&44 & [3] & $-$\\
IC~356 & 04~07~46.47 & $+$69~48~45.20 & SA(s)ab pec & 11&2   & 11&32 & [2] & F658N\,(A)\\
\n2146 & 06~18~37.71 & $+$78~21~25.30 & SB(s)ab pec & 17&2   & 11&04 & [4] & F658N\,(A), F160W\,(N)\\
\n2997 & 09~45~38.70 & $-$31~11~25.00 & SAB(rs)c    & 12&2   & 10&61 & [2] & F220W, F330W\,(A)\\
\n3756 & 11~36~47.97 & $+$54~17~37.25 & SAB(rs)bc   & 21&7   & 10&38 & [5] & F658N\,(A)\\
\n4041 & 12~02~12.17 & $+$62~08~14.20 & SA(rs)bc?   & 22&7   & 10&55 & [2] & F658N\,(A)\\
\n6217 & 16~32~39.22 & $+$78~11~53.60 & (R)SB(rs)bc & 18&3   & 10&46 & [2] & F658N\,(A)\\
\enddata
\tablecomments{Coordinates (J2000) correspond to the UVIS aperture positions. Morphologies are taken from \citet[][in the \ned\ homogenized notation]{rc3}, while distances are drawn from: [1]~\citet{mark3}; [2]~\citet{nbgc}; [3]~\citet{tully09}; [4]~\citet{kingfish}; [5]~\citet{springob09}. Stellar masses are derived through the `best' $K_S$ photometry from 2MASS \citep{2mass}. The final column (`Other \hst\ data') lists non-proprietary imaging datasets from ACS (A), WFPC2 (W2), and NICMOS (N) in the MAST archive using filters other than \bvi-equivalents. \n6217 was scheduled and observed with UVIS/F336W, but the observation failed.}
\end{deluxetable}
%% -------------------------------------------------------------------

The resulting sample of 22 galaxies was drawn from the combined target lists of the 11~Mpc \ha\ Ultraviolet Galaxy Survey~\citep[11HUGS;][]{11hugs}; the Local Volume Legacy survey~\citep[LVL;][]{lvl}; the  ACS Nearby Galaxy Survey Treasury~\citep[ANGST;][]{angst}; the  \hst\ \ha\ Snapshot survey~\citep[HHsnap;][]{hhsnap}; and the \cite{larsen04} catalog. The last catalog on that list contained size measurements, photometry and information on morphology for clusters in galaxies with WFPC2 imaging in various filters (typically B, V and I equivalents) and ground-based UBVRI imaging. Like the present work, the \citeauthor{larsen04} study was also aimed at studying star clusters and their immediate environment. 

Our Cycle 18 Snapshot program ``\hst\ U~band Survey of Star Clusters in Nearby Star-Forming Galaxies'' (PI~Smith, ID~12229) was awarded 22 snapshot orbits, one for each of 22 proposed targets. Eleven galaxies were observed, as listed in Table~\ref{tab:targ}, as per the nominal completion rate of 50\% for SNAP programs (one observation failed).  About half of observed galaxies are within $\approx12$~Mpc, and four systems lie at $18-23~$Mpc. All exposures were taken with the F336W filter and the duration was 1800~s. A three-point dither line pattern was chosen to cover the chip gap and to aid in the removal of hot pixels, cosmic rays and other artifacts. Either the UVIS-FIX or UVIS2-FIX aperture was used depending on the spatial extent of the galaxy compared to the 162\asec $\times$ 162\asec\ UVIS field of view. The precise pointings were chosen to give maximum overlap with the archival \bvi\ observations. 

Throughout this series we will use the Johnson filter notation with the specific \hst\ filter subscripted, but will not at any point convert between the two systems. For example, F336W will be denoted as \uub, while F555W and F606W, both roughly corresponding to Johnson $V$ band, will be written as $V_{555}$ and $V_{606}$. The \uub\ through \ib\ base will be referred to as \ubvi. 

% Subsection
% -----------------------------------------
\subsection{\n4041: Observations and Data Reduction}\label{sec:redu}

Our chosen proof-of-concept object was \n4041, a face-on SAbc galaxy near the upper end of our distance limit at 25~Mpc. This choice allows for a demonstration of the methodology of source selection and processing for approximately half the SHUCS galaxies, those at distances beyond $\approx15~$Mpc. The processing and analysis of nearby systems will be outlined in a future paper. In addition, it offers an interesting study of environmental effects on star cluster formation and evolution in a structurally segmented system. 

%% TABLE: OBSERVATIONS OF NGC 4041 -----------------------------------
\begin{deluxetable}{lclccl}
\tablewidth{0pt} 
\tablecolumns{6}
\tablecaption{Archival and new observations of \n4041\label{Tab:Obs}}
\tablehead{
\colhead {Instrument} & 
\colhead{Date} &
\colhead{Filter} &
\colhead{Exposure} &
\colhead {Program ID} &
\colhead{P.I. Name}\\
&&&
\colhead{(s)}
}
\startdata
WFC3/UVIS & 2011 Jan 30 & F336W & 1800 & 12229 & Smith \\
WFPC2 & 2001 Jul 04 & F450W & 320 & 9042 & Smartt \\
WFPC2 & 2001 Jul 04 & F606W & 320 & 9042 & Smartt \\
WFPC2 & 2001 Jul 04 & F814W & 320 & 9042 & Smartt \\
ACS/WFC & 2004 May 29 & F814W & 120 & 9788 & Ho \\
ACS/WFC & 2004 May 29 & F658N & 700 & 9788 & Ho \\
\enddata
\vspace{-5pt}\end{deluxetable}
%% -------------------------------------------------------------------

We discuss the structure and environment of this galaxy in Section~\ref{sec:n4041}  and show a composite \hst\ image in Figure~\ref{fig:hst}.
The \hst\ observations of \n4041 are shown in Table~\ref{Tab:Obs}. The archival \bb, \vb, \ib, and H$\alpha$ (F658N) data were obtained by two separate programs (9042 and 9788). All data were retrieved from the \textit{Mikulski Archive for Space Telescopes} (MAST) and each dataset was combined, corrected for geometric distortion, and drizzled to the native pixel scale using {\sc multidrizzle} \citep{multidrizzle_paper,multidrizzle}. For each WFPC2 image, two undithered exposures were obtained. These were drizzled separately for each CCD and the final pixel scale was $0\farcs05$ (PC) and $0\farcs10$ (WF2, 3, 4). The ACS F658N data consisted of a pair of images and these were drizzled to a pixel scale of $0\farcs05$. Only a single F814W ACS image was taken; cosmic rays  were removed using the {\sc lacosmic} task \citep{lacosmic} and the resultant image drizzled. For the WFC3/UVIS data, three dithered images were obtained and these were drizzled to a pixel scale of $0\farcs04$.

We have not corrected the images or photometry for charge transfer efficiency (CTE) degradation. The ACS images were taken two years after the instrument was installed and CTE losses at this epoch are negligible \citep{ubeda12isr}. CTE losses are apparent in the individual WFC3/UVIS images but the background is above the critical level of 5 electrons where CTE losses are severe \citep[see][]{noeske12isr}. Given that there was no correction available when we processed the WFC3 data, and the CTE losses are a small effect, we have not taken CTE losses into account. For WFPC2 data, no CTE trails are apparent in the images because of the fairly high backgrounds and we have therefore not applied any photometric corrections to the WFPC2 data.

% =============================================
% 3. CLUSTER SELECTION AND PHOTOMETRY PIPELINE
% =============================================
\section{Selection of Star Cluster Candidates and Data Products}\label{sec:pipeline}
\noindent
Star clusters are only marginally resolved at the distance to \n4041, as the typical diameter of 7~pc subtends an angle of $0\farcs1$ on the sky. This translates to one or two pixels on various \hst\ cameras. A considerable drawback is hence the potential inclusion of stellar associations, which are not discriminated by automatic source detection algorithms~\citep[see][]{silvavilla10,bastian11}. Toward that end we employ the \textit{concentration index} (CI), defined as the difference in brightness between two apertures: one comparable to the size of the point-spread function (PSF), and another representing the typical star cluster \citep[see][]{whitmore07}. This technique helps to place clusters between stars, which have very small CI, and associations, at large CI. We will develop the use of this method in and Section~\ref{sec:CI}.

With the above in mind, we developed a selection and photometry pipeline, which is run entirely in IRAF\footnote{
	IRAF is distributed by the National Optical 
	Astronomy Observatories, which are operated 
	by the Association of Universities for Research
	in Astronomy, Inc., under cooperative agreement 
	with the National Science Foundation.} 
and consists of the following steps: 
\begin{enumerate}%[topsep=2pt,itemsep=0pt,parsep=3pt]

%%% STEP 1: 
\item {\bf Source selection in \uub}\label{pipeline:step1}. We run \texttt{DAOfind} to select sources brighter than $7\,\sigma$ above background, measured on various parts of the image. We do not restrict `roundness' (axial ratio) or `sharpness' (size compared to the stellar full width at half-maximum), in order to include elliptical and marginally resolved clusters (the majority at this distance, as noted above). 

%%% STEP 2: 
\item {\bf Coordinate transformation}. Our dataset consists of images taken for different projects at various times and with different cameras, which often leads to overlapping, but not identical pointings. Our \uub\ observations were designed for maximum overlap with the archival imaging, and hence serve as the reference frame for the World Coordinate System (WCS). The coordinate lists from Step~\ref{pipeline:step1} are converted from the \uub\ frame to those of each instrument/pointing used in the study. This step uses the \texttt{tmatch} algorithm with typically some 20 reference sources {(stars or compact clusters) that span the entire image as much as possible}. The process is refined until the root-mean-squared errors for x and y positions do not exceed 0.1~px, a scale tested to eliminate source confusion. When registering WFPC2 data, the four CCDs are mapped individually, and the number of reference stars is usually ten or lower.\label{pipeline:step2}

%%% STEP 3: 
\item {\bf Multi-band photometry}. Photometric apertures of radius 0\farcs12 are placed at the coordinates defined in Step~\ref{pipeline:step2}, with the background measured locally in annuli of 0\farcs2 to 0\farcs3, depending on the pixel scale of each camera and the corresponding PSF. These values were derived after testing the typical growth curves of modeled stellar PSFs generated with {\tt Tinytim} \citep{tinytim}, coupled with the \textit{STScI Focus Model Utility}\footnote{http://www.stsci.edu/hst/observatory/focus/FocusModel} to account for breathing at the time of observation. We note that these values will be revised for each galaxy depending on its distance. We do not allow the \texttt{DAOphot} task to re-center sources as the differing pixel scale between cameras introduces non-negligible positional uncertainties. Given the employed pixel scales of the WFC3/UVIS (0\farcs040~px$^{-1}$), ACS/WFC (0\farcs050~px$^{-1}$), WFPC2/PC (0\farcs045~px$^{-1}$), WFPC2/WF (0\farcs100~px$^{-1}$), the radius of the applied photometric aperture corresponds to image sizes of 3.0, 2.4, 2.7, and 1.2~px respectively. In the test case of \n4041, these translate to a physical radius of 13.4~pc, \cf\ the typical effective cluster radius of $\sim3.5~$pc. The photometric error is computed according to the \texttt{DAOphot} recipe that takes into account the gain and readout noise of each detector. For cases where imaging from more than one \hst\ camera is available in a given filter, we photometer all images and choose the measurement with the lowest associated error. We calibrate the photometry in the Vegamag system. 

The photometric process is completed by the measurement of a CI for each source as the difference in \uub\ brightness between apertures of one and three pixels. Such a technique has been demonstrated in the past to produce viable samples of star clusters at comparable distances \citep[\eg\ The Antenn\ae,][]{whitmore07}. More information can be found in Section~\ref{sec:CI}. \label{pipeline:step3}

%%% STEP 4: 
\item {\bf Photometric corrections}. At this stage we correct for the size of the aperture. This has to be tailored to each galaxy individually, as the correction is dependent on distance and whether the sources are resolved or not. In addition, we do not always have enough bright, isolated clusters to derive growth curves.  The corrections thus combine our own empirical deductions with the encircled energy curves given for each instrument. For the test case of \n4041, isolated sources were very rare, so corrections were derived through the growth curves of model PSFs, generated for each detector-filter combination with {\tt Tinytim}. 
Finally, we add to our photometric errors a factor of 0.05\,mag in quadrature, to account for uncertainties in the photometric zeropoints \citep[for details see][]{adamo10a}, and correct for \mbox{foreground extinction \citep{schlegel98}}.\label{pipeline:step4}
\end{enumerate}%\vspace{-15pt}

% Subsection
% -----------------------------------------
\subsection{Completeness estimation}\label{sec:completeness}
At the distance to \n4041, the bulk of its cluster population is nearly point-like. Therefore, we follow the traditional method of artificial star counts to estimate completeness. In future works focussed on nearby systems, however, we will use artificial clusters instead, to overcome the uncertainties introduced by stochastic variations of the colors of low-mass clusters \citep[introduced by under-sampled stellar mass functions; see][]{fouesneau10,popescu10,silvavilla11}. 

% Figure: Completeness
% ------------------------
\begin{figure}[htb]
	\begin{center}

		\includegraphics[width=0.48\textwidth,angle=0]{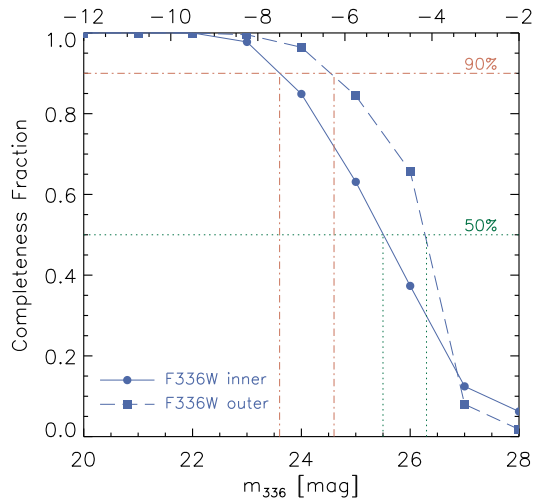}
		\caption{Completeness tests tracing the fraction 
			of artificial stars recovered at a range of 
			brightness. The dotted vertical lines indicate 
			50\% and 90\% completeness fractions for the 
			inner disk and outer galaxy, performed 
			separately to account for the different background
			levels. The top axis displays the range of 
			absolute magnitudes covered. We estimate  50\% 
			and 90\% completeness levels at (25.5, 23.6)~mag 
			and (26.3, 24.6)~mag for the inner disk and 
			outer galaxy respectively. 
		}\label{fig:completeness}

	\end{center}
\end{figure}
% ------------------------

We created sets of artificial stars in a $15\times15$ object grid, using the \texttt{MKSynth} algorithm of \citet{ishape}. We used \texttt{TinyTim} models of the WFC3 PSF as described in Step~(\ref{pipeline:step3}) of the pipeline (Section~\ref{sec:pipeline}). The generated objects were assigned magnitudes in the range [20, 30]~mag, in steps of one magnitude. Since completeness varies with the local background \citep[\eg][]{remco07}, as well as crowding and confusion, we use two fields for this process: one covering most of the inner disk, and another, equally-sized field that covers the outer spiral structure. Results are presented in Fig.~\ref{fig:completeness}, where we show the 50\% and 90\% recovery fractions to occur at (25.5, 23.6)~mag and (26.3, 24.6)~mag for the inner disk and outer galaxy respectively. 

% Figure: brightness-error
% ------------------------
\begin{figure}[tbhp]

	\begin{center}

		\includegraphics[width=0.47\textwidth,angle=0]{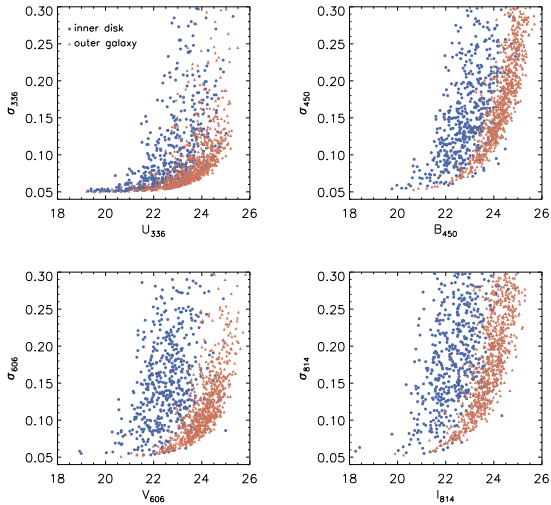}
		\caption{Brightness-error plots across the \ubvi\ 
			filter-set. The separation into inner and outer galaxy 
			(blue dots, red triangles) accentuates the issues faced 
			with photometric errors in the inner disk, an effect of 
			crowding. This is taken into account in all analysis. 
			Using this plot we tentatively establish \vb\ as the 
			limiting band for this study, as it shows the most 
			irregular brightness-error distribution. Regular 
			statistical measures, such as the standard deviation of 
			the error distribution, provide no distinction between 
			the \bvi\ filters.}
			\label{fig:err4}

	\end{center}
\end{figure}
% ------------------------

Another photometric property that will impact our star cluster analysis is the limiting filter. We estimate this through the brightness-error plot of Figure~\ref{fig:err4}. Regular statistical measures, such as the standard deviation of the error distribution, provide no distinction between the \bvi\ filters. We therefore look for the most irregular distribution and adopt \vb\ as the limiting band.

% Subsection
% -----------------------------------------
\subsection{Concentration Index}\label{sec:CI}
In Step~\ref{pipeline:step3} of the photometry pipeline (Section~\ref{sec:pipeline}) we note the measurement of the CI, {as the difference in magnitude between apertures of radius 1 and 3 px. We employ this metric in order to distinguish between star clusters and stellar associations, large, short-lived structures \citep[see][]{pz10,bastian11,bastian12} that can contaminate samples and affect the analysis of star cluster populations.}

\begin{figure}[tbhp]

	\begin{center}

		\includegraphics[width=0.5\textwidth,angle=0]{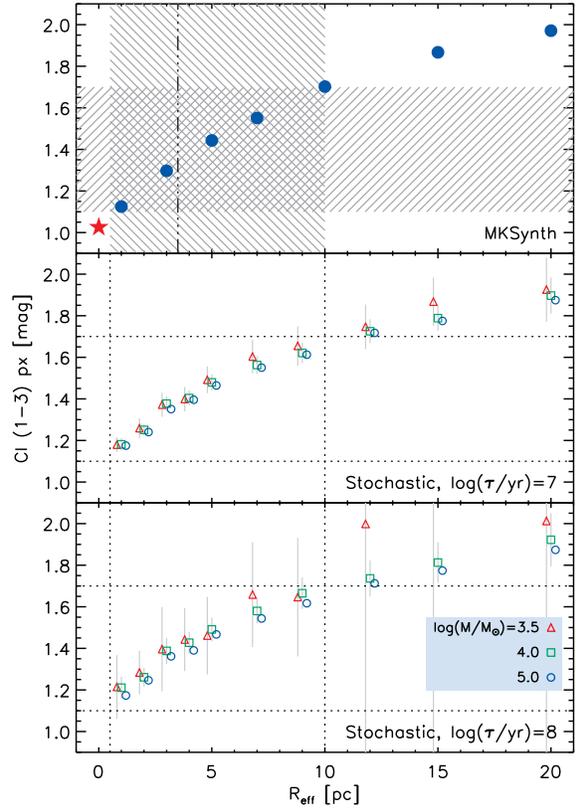}
		\caption{Concentration Index tests on artificial 
			clusters. 
			The {\bf top} panel shows a CI versus radius 
			plot for clusters created as \citet{eff87}
			light profiles of index $\gamma=3.0$, convolved 
			with an empirical PSF. The {\bf center} and 
			{\bf bottom} panels show artificial clusters 
			created as aggregates of individual stars, 
			arranged according to a King profile (more 
			appropriate for older clusters). Cluster ages 
			are $\log(\tau/\textup{yr}) = 7, 8$, while 
			different symbols represent a range of masses. 
			As they age, clusters display less regular 
			behavior, as evidenced by the error bars in 
			the $\log(\tau/\textup{yr}) = 8$ model set. 
			A consistent asymptotic behavior is noted 
			through both methods. $\reffmath\gtrsim10~$pc 
			are rarely observed, so we set the CI at this 
			radius as the upper limit. An $\reffmath=0$ 
			represents individual stars, and the corresponding 
			CI$~\simeq1.0$~mag is adopted as the lower 
			limit. 
		}\label{fig:CImodel}

	\end{center}
\end{figure}

In order to employ a CI cut in our source filtering process, we test the theoretical expectation from modeled clusters of various ages and sizes. First we use \texttt{MKSynth} (see Section~\ref{sec:completeness}) to convolve an \citet{eff87} surface brightness profile with a PSF derived from point sources in our images. Since we could not identify suitable stars on the F336W image of \n4041, we employed SHUCS images of \n891 and \n2146. In addition, while various images are taken at different focus settings, the PSF does not change to a level that will affect our analysis -- the effect is smaller than our typical photometric error of 0.1~mag. We assign a range of sizes to these model clusters, incorporate them on the F336W image and measure their CI in the same way as for the detected sources. The top panel of Figure~\ref{fig:CImodel} shows a plot of CI versus assigned radius. The first datapoint, at $\reffmath=0~$pc, denotes individual stars, which register values of $\approx1.0~$mag. We therefore adopt a lower limit for cluster candidates at $\textup{CI}=1.1~$mag. We adopt an upper limit from the literature, as the CI of the largest observed clusters of approximately 10~pc \citep[see size distributions of][]{larsen04,barmby06,remco07,mayya08,bastian12}. The corresponding CI is 1.8~mag on Figure~\ref{fig:CImodel}. 

To confirm the asymptotic shape of this relation we employ clusters generated as aggregates of individually synthesized stars, following the methodology of \citet{silvavilla10}. Each star is represented by the empirical PSF described in Section~\ref{sec:completeness} and a King profile of index $c=\frac{r_\textup{\scriptsize tidal}}{r_\textup{\scriptsize core}}=30$, while the age, mass and effective (half-light) radius assume the following ranges: 

$$ \log(M/\Msun) = [3.0, 3.5, 4.0, 4.5, 5.0] $$
$$ \log(\tau/yr) = [6.6, 7.0, 8.0, 9.0] $$  
$$ \reffmath = [1,2,3,4,5,7,9, 12,15,20]~\textup{pc} $$

The range of CI assigned to clusters is shown as a hashed area in the top panel of Figure~\ref{fig:CImodel}, while in the center and bottom panels, the CI area is bracketed by dotted lines to aid illustration. The CIs of stochastically modeled clusters are slightly higher (by $\approx0.1~$mag), owing to their King surface brightness profiles, chosen to generate more realistic older clusters. 

% Figure: model CI vs radius
% ------------------------
% ------------------------

\begin{figure}[!h]

	\begin{center}

		\includegraphics[width=0.48\textwidth,angle=0]{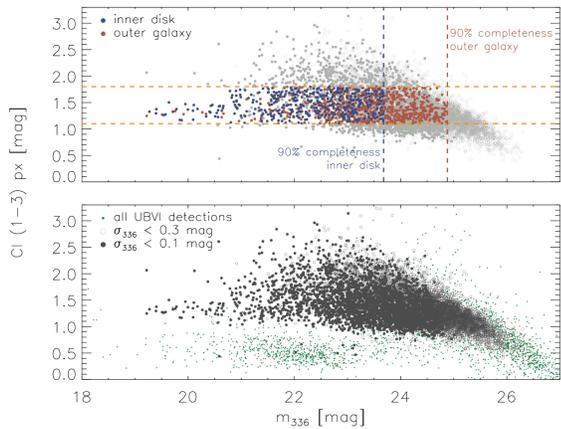}
		\caption{Concentration Index versus source brightness
			on isotropic axes. 
			The {\bf bottom} panel shows all sources that 
			register \ubvi\ photometry as green points, while 
			the open and closed gray circles apply a cut 
			according to photometric error. 
			The {\bf top} panel shows the space chosen to 
			represent high-quality cluster candidates, the 
			ones used for all further analysis, against a 
			backdrop of the gray circles of the lower panel. 
			Vertical lines denote the 90\% completeness limits
			for the inner and outer galaxy. 
		}\label{fig:CI}

	\end{center}
\end{figure}

The measured CI values of actual sources are plotted against \uub\ brightness in Figure~\ref{fig:CI}. Applying the above CI cut and the 90\% completeness limits of Section~\ref{sec:completeness}, we restrict cluster candidates to a very specific part of photometric parameter space. 

% Figure: CI vs brightness
% ------------------------
% ------------------------

% Subsection
% -----------------------------------------
\subsection{SED-fitting with \ygg\ models: derivation of age, mass and extinction}\label{sec:sed}
\begin{figure*}[tbhp]

	\begin{center}

		\includegraphics[width=\textwidth,angle=0]{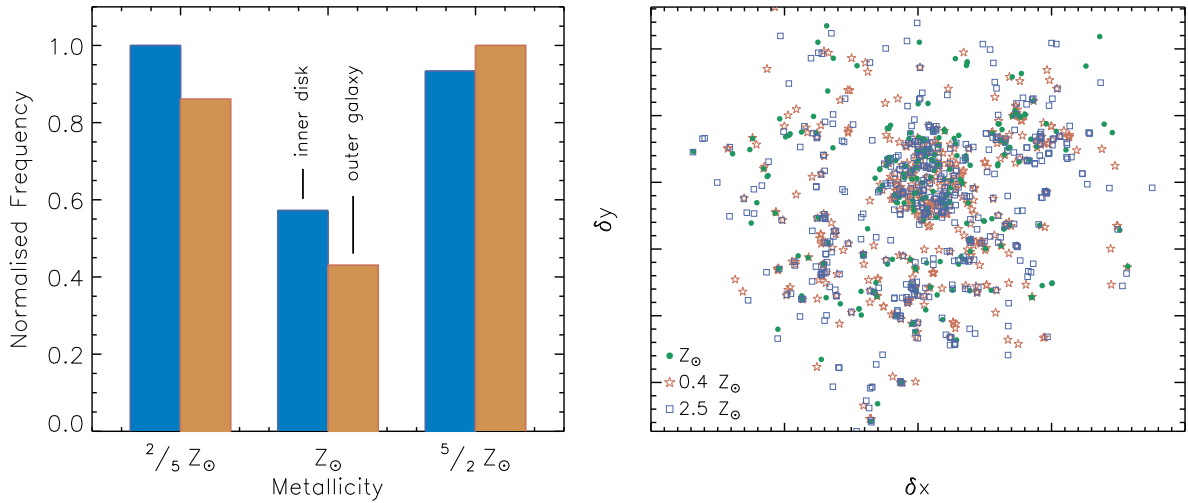}
		\caption{Metallicity distribution ({\bf left}) of all 
			sources in the filtered catalogue (\fubvi~$=1111$, 
			see Section~\ref{sec:data-products}), divided according
			to galactocentric radius. We find sub- 
			and super-solar metallicities to be preferred in 
			\n4041. However, the uncertainties in photometrically 
			derived metallicity are considerable, so no firm 
			conclusions can be drawn from this histogram. 
			The metallicity distribution also does not show 
			any spatial trends. This is shown on the {\bf right},
			where all clusters in the final sample are symbol 
			coded according to derived metal content.  
		}\label{fig:metal}

	\end{center}
\end{figure*}

The age, mass, and extinction of all candidate star clusters are derived through fits to their spectral energy distribution (SED) with \ygg\ models \citep[see][for a detailed description]{zackrisson11}. Such a model set is ideal for our work, as it combines single stellar population (SSP) synthesis models with nebular spectra, both emission lines and continuum, and is tuned to the redshift of each galaxy. Analyses using \ygg\ models have in the past been successful in describing the cluster populations of intensely star forming galaxies \citep[\eg][]{adamo10b}. For the SHUCS sample, we use Starburst99 SSPs \citep{sb99,vazquez05}, compiled using a \citet{kroupa01} stellar initial mass function throughout the mass range \mbox{$0.1-100~$\Msun}; Padova stellar evolutionary tracks; and three different metallicities: 0.4, 1, and 2.5 times solar. We note no spatial trends in terms of metallicity, as shown in Figure~\ref{fig:metal}. While this result might seem to oppose the well-established metallicity gradients in local spirals \citep[\eg\ in the Local Group;][]{cioni09}, the photometric derivation of metallicity is subject to large uncertainties. No firm conclusions can be drawn about the metallicity distribution. To reproduce the integrated fluxes of the earlier stages of cluster evolution, \ie\ luminous H\two\ regions surrounding the recently formed stars, the models include a self-consistent treatment of the nebular emission performed with the Cloudy photoionization code \citep{cloudy}. The metallicities of the gas and the stars are assumed to be the same.  Using the gas properties of typical local H\two\ regions \citep{kewley02} we assume a filling factor $f= 0.01$, a hydrogen density of $n(\textup{H}) = 100~$cm$^{-3}$, and a covering factor, \ie\ fraction of ionizing photons absorbed by the gas, of $c = 0.5$. Different values of $c$, $n(\textup{H})$, and  $f$ only impact the SED fits for the very youngest clusters. We therefore never attempt to resolve the $1-5~$Myr age block, and accept a single, `young' age for such sources. 

% Figure: Metallicity
% ------------------------
% ------------------------

These models, together with the photometric catalogs, comprise the input for a least-$\chi^2$ fitting algorithm, as detailed  in \citet{adamo10a}. In brief, the program considers sources detected in four or more filters and applies the \citet{cardelli89} attenuation law at each age step to constrain the visual extinction in the line of sight. The final age, mass, and extinction are assigned to a cluster given the set of parameters that minimizes the $\chi^2$. Uncertainties are carried through the fitting process. The program estimates the best reduced $\chi^2_{\nu,best}$ ($\chi^2$ divided by the degrees of freedom), and saves, as the range of acceptable values, all the solutions with $\chi^2 \leq \chi^2_{\nu,best}+3.5$. The maximum and minimum age, mass, and extinction contained in this pool of likely solutions are assigned as errors. This is a typical method of estimating errors when the $\chi^2$ statistic is employed, \eg\ the photometric fits of \citet{bik03} and the spectroscopic fits of \citet{isk09a}. \citet{lampton76} provide a full reasoning. 

\begin{figure*}[htbp]

	\includegraphics[scale=0.42]{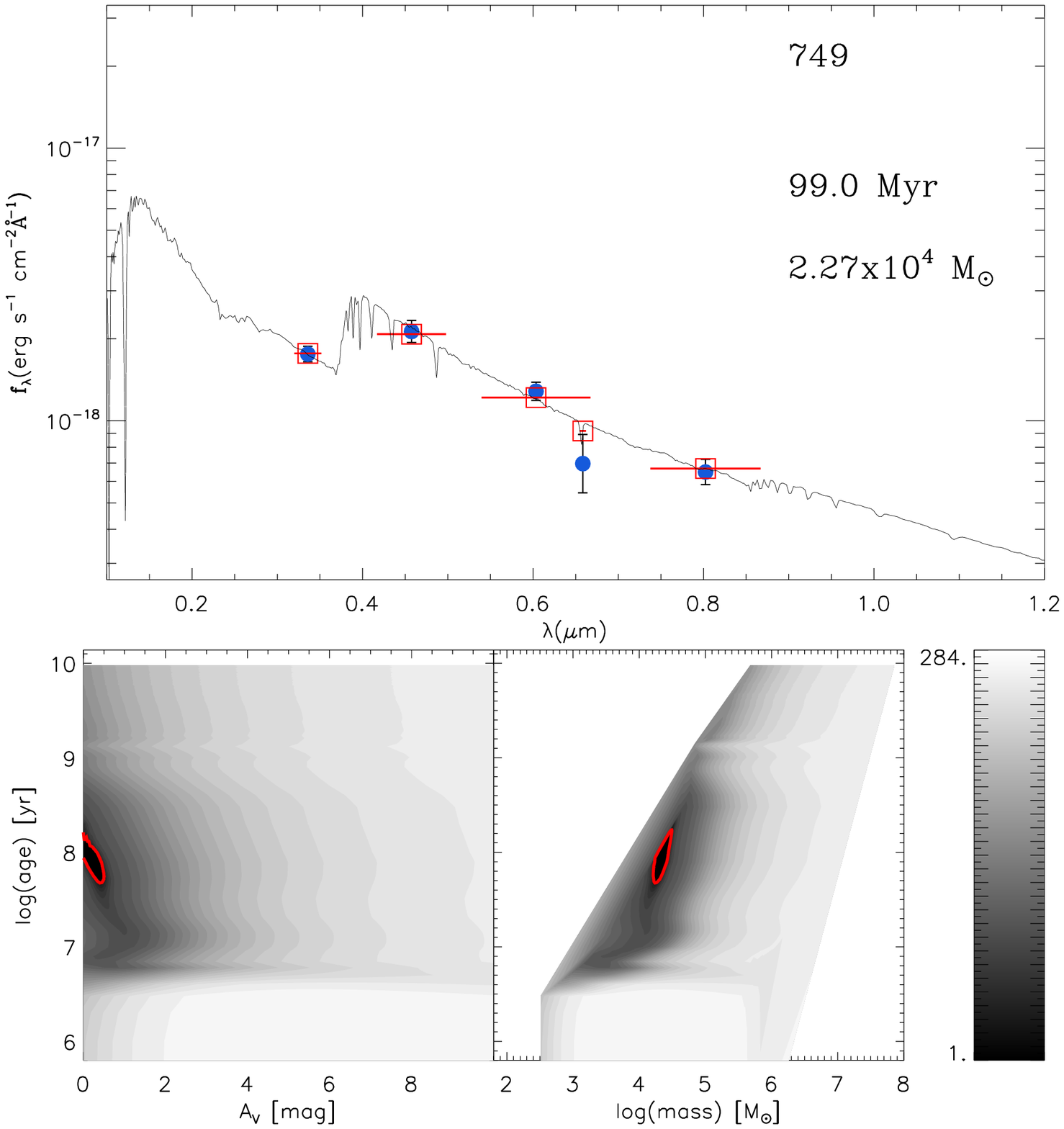}
	\includegraphics[scale=0.42]{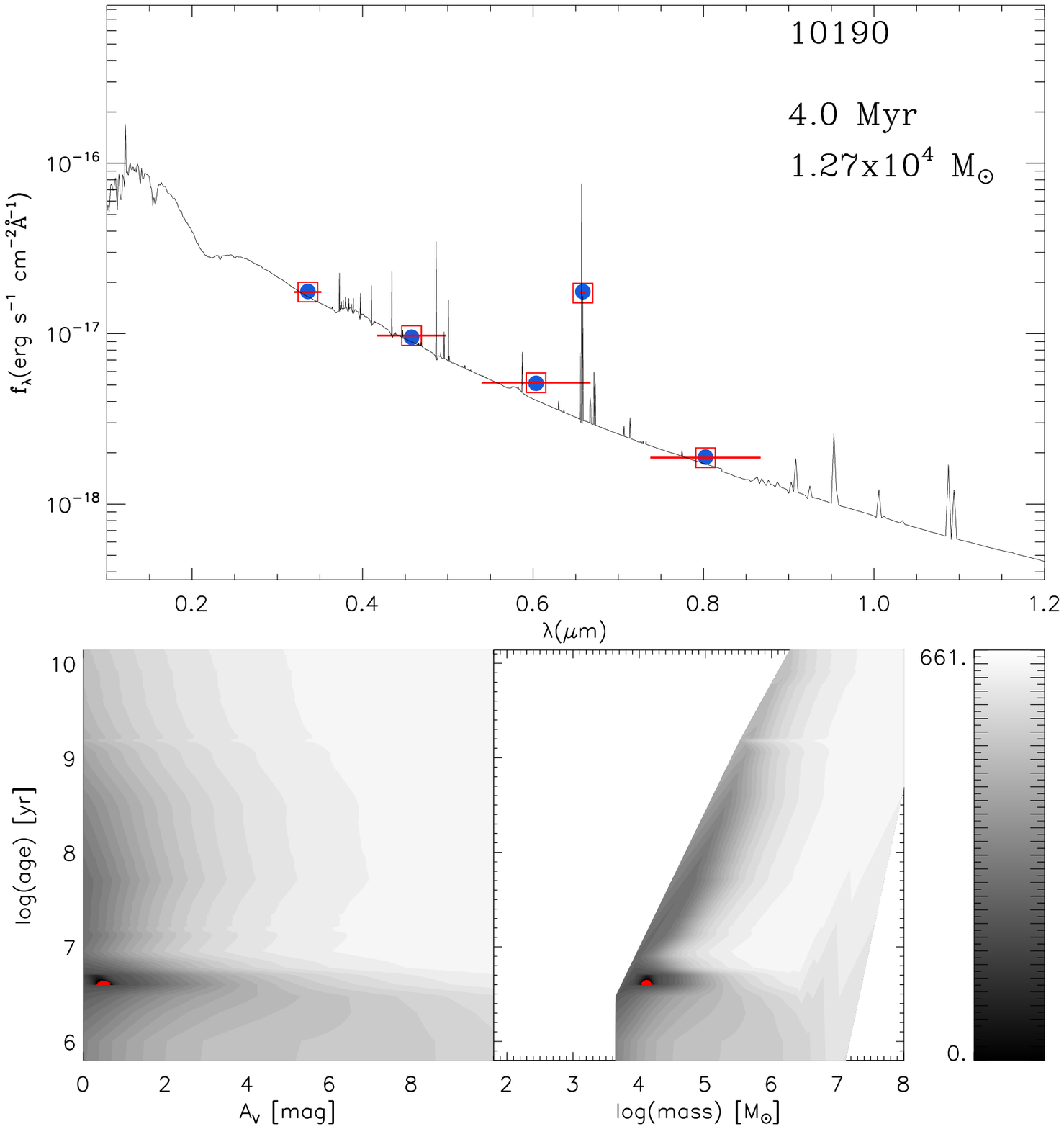}\\
	\includegraphics[scale=0.42]{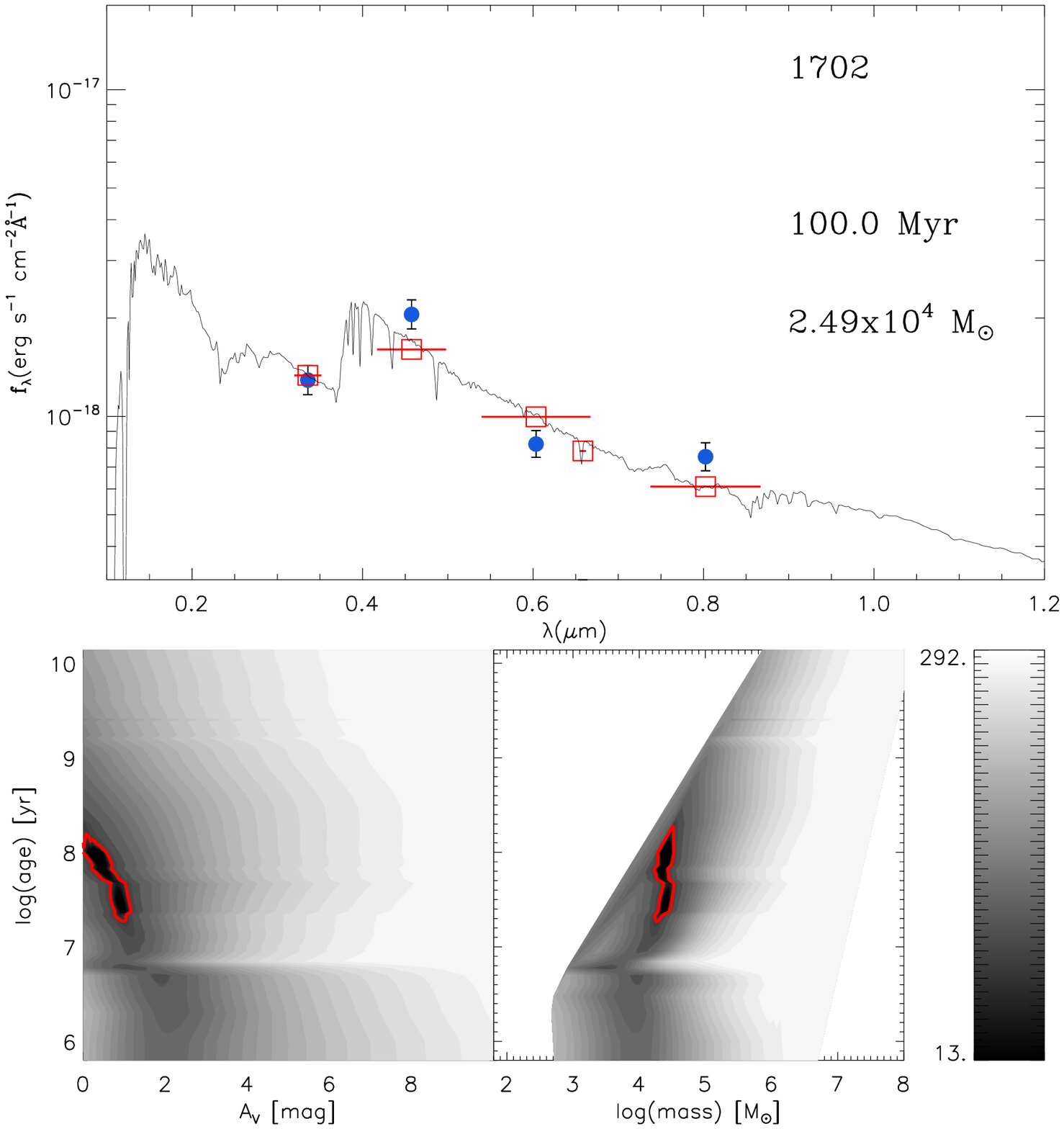}
	\includegraphics[scale=0.42]{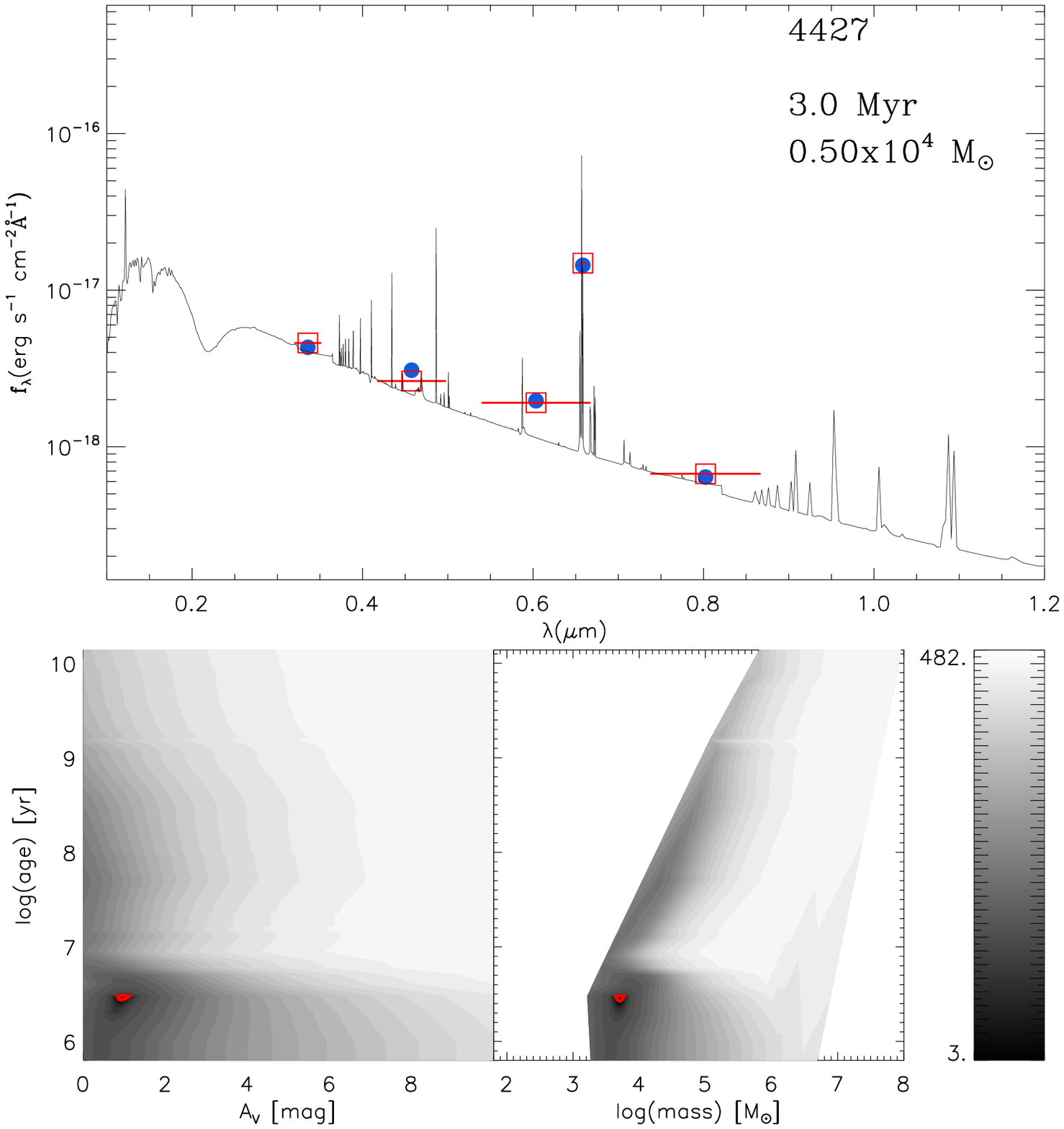}

	\caption{SED fits of four cluster candidates in \n4041 
		with {\it Yggdrasil} models, following the process 
		of \citet{adamo10a}. The filled blue dots represent 
		the observed integrated fluxes, with associated error 
		bars. The corresponding integrated model fluxes are 
		labelled as red squares and the horizontal lines 
		indicate the bandpass of the given \hst\ filter. 
		The quality of the fit is shown in the lower panels 
		of each set, where the $\chi^2_\nu$ is plotted across the 
		parameter spaces of age-extinction and age-mass.  
		The red contour line indicates the range of acceptable 
		fit solutions. Each plot is labelled with an identifier, 
		and the derived age and mass of the cluster candidate. 
		The top row shows good fits, while the cluster 
		candidates of the bottom row are filtered out of the 
		final catalog. Cluster 4427 is found in a crowded 
		region and its light is contaminated by neighboring 
		sources. As such it does not pass the CI 
		test employed by the filtering process and receives 
		\fubvi~$=11$. 
	}
\label{fig:sed_fits}
\end{figure*}

Example SED fits are shown in Figure~\ref{fig:sed_fits}. First we provide an illustration of the best fitting \ygg\ model, with source photometry marked as filled dots. The red squares mark the best-fit model flux on the y axis, along with the central wavelength and bandpass of each employed \hst\ filter. The plots of the bottom row plot $\chi^2$ in the age-extinction and age-mass spaces. Red contours mark the range of acceptable solutions to the SED fit, defined as $\chi^2_\nu < \chi^2_{\nu,best}+\frac{3.5}{\nu}$. The top row of figure sets features two cluster candidates that meet the selection criteria (detailed in next section), while the ones in the bottom row do not. 

% Figure: Example SED fits
% ------------------------

% Subsection
% -----------------------------------------
\subsection{Cluster candidate selection and data products}\label{sec:data-products}
The primary data products of each study are a Photometry Table and an SED Table. These data products will be made available to the community in their entirety when the survey has been completed. The Photometry Table collects positional and photometric measurements and applies a photometric flag, \fubvi, according to the following scheme: 

%~\phantom{111}1:~Source detected and photometered across\\ 
%$\quad\quad$the \ubvi\ baseline,\\
%\indent~\phantom{11}11:~\ldots plus \ubvi\ errors less than 0.3~mag,\\
%\indent~\phantom{1}111:~\ldots plus a CI in the range [1.1, 1.8]~mag, \\
%\indent~1111:~\ldots plus \uub\ brighter than the 90\% completeness limit.
\begin{itemize}[topsep=0pt, itemsep=0pt, parsep=0pt]
	\setitemize[0]{leftmargin=5pt}%,itemindent=10pt}
	\item [1:]    Source detected and photometered in \ubvi, 
	\item [11:]   \ldots plus \ubvi\ errors less than 0.3~mag, 
	\item [111:]  \ldots plus a CI in the range [1.1, 1.8]~mag,
	\item [1111:] \ldots plus \uub\ brighter than the 90\% completeness limit.
\end{itemize}

All analysis is performed on clusters with \fubvi~$=1111$. The scheme is complemented by a second flag, \fubvi$_+$, which also considers the photometric quality in bands other than \ubvi~(in this case \ha), and is marked with a similar sequence of the number 2 for clarity (2, 22, 222, 2222). The effect of filtering the sample across the \ubvi\ baseline on photometric completeness is negligible, and will be discussed fully in the context of the luminosity functions of Section~\ref{sec:sc:LF}.

\begin{figure*}[!t]

	\begin{center}

		\includegraphics[width=\textwidth,angle=0]{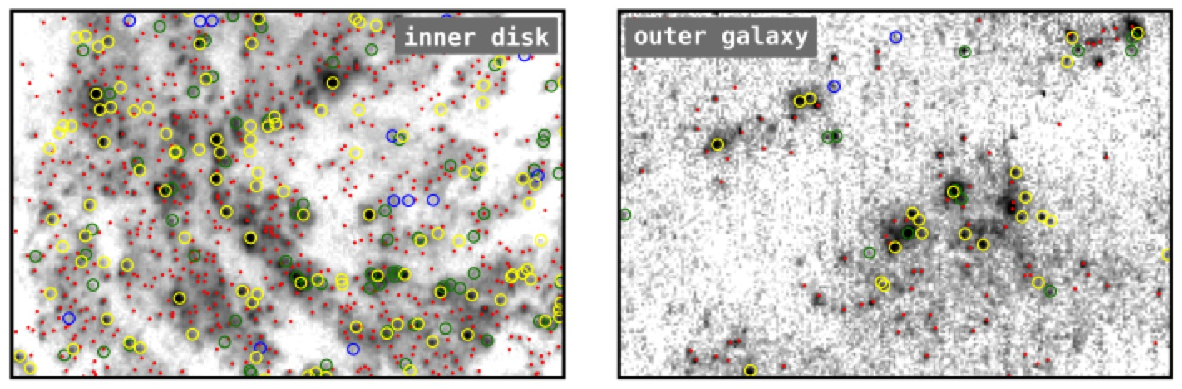}
		\includegraphics[width=\textwidth,angle=0]{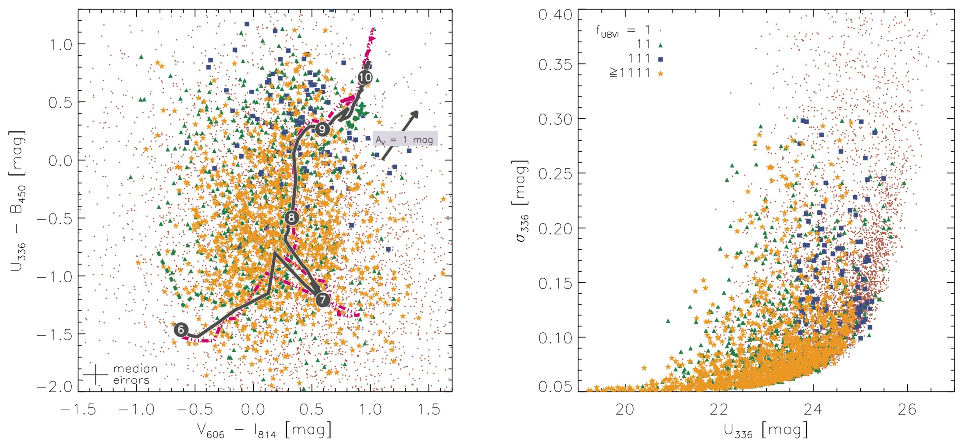}
		\caption{The cluster selection process, demonstrated 
		for the population of \n4041. 
		The {\bf top} panels show close-ups of the F336W frame 
		(inner, outer galaxy on the left and right). Each image 
		measures at approximately $1.3\times0.9~$kpc. We mark 
		\fubvi\ values of 1, 11, 111, and 1111 as red dots and 
		green, blue, and yellow circles of radius 3~px, coinciding 
		with the photometric aperture. 
		The {\bf bottom} panels follow this selection on a 
		\ubvi\ color diagram, with associated single stellar
		population model tracks \citep{zackrisson11}. The 
		metallicities of the dashed (green), solid (black) and 
		dashed-dotted (magenta) lines are $0.4, 1.0, 2.4\times$\Zsun. 
		On the right we show a \uub\ magnitude-error plot. None 
		of the criteria appear to introduce color-space biases.
		}\label{fig:select}

	\end{center}
\end{figure*}

This filtering process, as applies to \n4041, is demonstrated in Fig.~\ref{fig:select}. The top row shows two segments of our F336W image where we mark sources with \fubvi\ of 1 through 1111 as red dots and green, blue, and yellow circles. The radius of 3~px matches the photometric aperture in \uub. The bottom row shows a \ubvi\ color diagram, color-coded in the same way, to demonstrate the effect of each filter on the selection process, along with the corresponding brightness-error plot. From these images it is clear that the distance to this galaxy gives rise to a crowded field of clusters and associations. Coupled with the stochastic variations in the light profile (present at all distances), this severely cuts down the number of clusters for which reliable size measurements can be performed. At $22.7~$Mpc one UVIS pixel corresponds to $\approx4~$pc, while one WFPC2/WF pixel is $11~$pc. PSF deconvolution codes \citep[\eg\ IShape;][]{ishape} have been demonstrated to resolve features as small as $0.4~$px on \hst\ imaging, which corresponds to just under $2~$pc. That way, the smallest effective radius that can be sampled is $\approx1~$pc. It is doubtful that such a source would have a sufficiently high signal-to-noise ratio to be detected, however, as a test run of IShape on \n4041 fit only large clusters, typically $\reffmath > 5~$pc. This topic will be revisited in future SHUCS works that treat nearby systems.

% Figure: source selection
% ------------------------
% ------------------------

In addition to the photometry flags, the Photometry Table also marks each source with an integer between 1 and 4 to denote which WFPC2 CCD covers the source. A value of `9' marks a detector other than WFPC2, while `0' indicates that a source is covered by WFPC2, but it is too close to the edge of the image. This flag is then repeated for each filter in the analysis, standardized as U, B, V, H$\alpha$, I, J (\eg\ NICMOS-F110W), H (\eg\ NICMOS-F110W). The shorthand filter representations are defined in the table header. Positional information is also included, in the form of coordinates measured on the F336W image. 

The results of the SED fits are collected in an SED Table, featuring a minimum, best-fit, and maximum value for age, mass, and extinction; the number of data points (bands) employed by the fit; and the reduced $\chi^2$ statistic of each fit.

%======= 4. NGC 4041: STRUCTURE AND ENVIRONMENT =======
\section{The Structure and Environment of \n4041}\label{sec:n4041}
We begin the analysis of the SHUCS dataset with the SA(rs)bc galaxy \n4041 \citep{rc3}. We adopt a distance of 22.7~Mpc and a corresponding distance modulus of 31.78 from \citet{nbgc}. Given this distance modulus, we derive an absolute magnitude of $g = -19.7~$mag, according to its Sloan Digital Sky Survey Data Release 6 \citep[SDSS;][]{sdss,sdss_dr6} model or Petrosian radius brightness. 

In Figure~\ref{fig:hst} we showed an \hst\ color-composite image of the galaxy in \ubvi\ and H$\alpha$. The galaxy appears to have a partitioned structure, with tightly wound inner spiral arms giving way to a looser, outer network. The boundary is defined by a smooth, yellow bulge-like feature, suggestive of an old underlying stellar population. The bulge is overlaid with bright, blue star clusters and dust lanes, and it fades quite rapidly toward the outskirts of the galaxy, where blue star clusters can be found in large numbers. Finally, we observe the nuclear source to be split into two components with comparable fluxes, both offset from the center of the outermost isophote by $0\farcs8$ and $1^{\prime\prime}$. The following sections treat the galaxy and its surroundings in more detail.

% Subsection: n4041 SBP
% -----------------------------------------
\subsection{A complex isophotal structure}\label{sec:sbp}
Multi-component spirals have been observed in the past, first the southern supergiant \n6902 \citep{jsg79}, and then in the context of low surface brightness galaxies \citep{bosma93}. \citet{matthews97} observed such structures in a sub-sample of `extreme' late-type galaxies. They also find offset central peaks in such galaxies, as do \citet{odewahn96} in another catalog of late-types, and \citet{karachentsev93} in Magellanic spirals. \citeauthor{matthews97} discuss the possibility of offset central peaks representing the true dynamical centers of their hosts, \ie\ centers of gravity offset from the geometric or isophotal center. \citet{marconi03} touch on this scenario by suggesting that the dynamical center of \n4041 might be decoupled from its geometric center. 

% Figure: n4041 SBP
% ------------------------
\begin{figure}[hpbt]
	\begin{center}

		\includegraphics[width=0.47\textwidth]{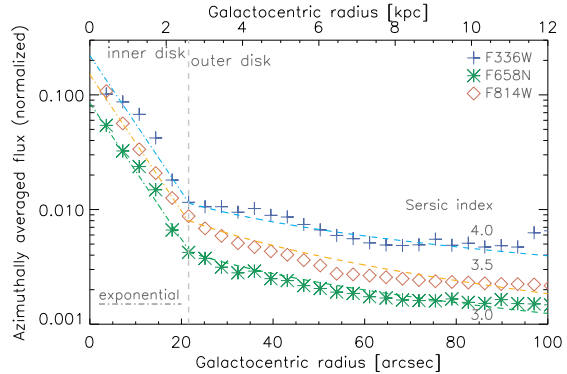}
		\caption{%
		Peak-normalised, azimuthally-averaged count rate of \n4041 
		in bins of $3\farcs5$ ($400~$pc) about the central peak. 
		While the inner disk is adequately described by exponential 
		flux profiles, the outer structure follows a profile more 
		suited to early-type galaxies -- a S\'ersic index of $4.0$ 
		is equivalent to the \mbox{\citet{devauc48}} $r^{1/4}$ law, 
		while an index of 1 equals an exponential profile. 
		}\label{fig:sbp}

	\end{center}
\end{figure}

The composite structure of \n4041 is investigated further through plotting the surface brightness of the galaxy in azimuthally averaged bins, normalized to the central peak (despite the small central offset, to simplify illustration). This is shown in Figure~\ref{fig:sbp}, where the central exponential disk measures at a radius of $\approx22^{\prime\prime}$ (2.4~kpc) and  we derive a radius of $\approx80^{\prime\prime}$ (9~kpc) for the outer galaxy from the optical isophotes. As noted above, the spiral structure is not discontinuous, with {some spiral arms stemming from the flocculent inner structure and developing into} loose outer arms. Across the disk the inter-arm distance increases from $\sim0.7~$kpc to $\approx1.1~$kpc, as measured on the F336W image, and the arm thickness is $\approx0.5~$kpc. The inner disk follows an exponential profile, as is expected for the disks of spiral galaxies, while the outer disk is better described by \citet{sersic68} profiles of high index, typical of early-type galaxies \citep[\eg][]{cote06}. The F336W profile shows a periodic fluctuation. While one might expect the population of young star clusters to contribute to this, \citet{larsen04} show the contrary: star clusters are never found to contribute more than 10\% of the overall U band luminosity of a spiral galaxy (and up to 20\% in starbursts).

% Subsection: n4041 group
% -----------------------------------------
\subsection{Neighbors and Potential Companions}\label{sec:group}
\begin{figure*}[!t]
	\begin{center}

		\includegraphics[width=\textwidth]{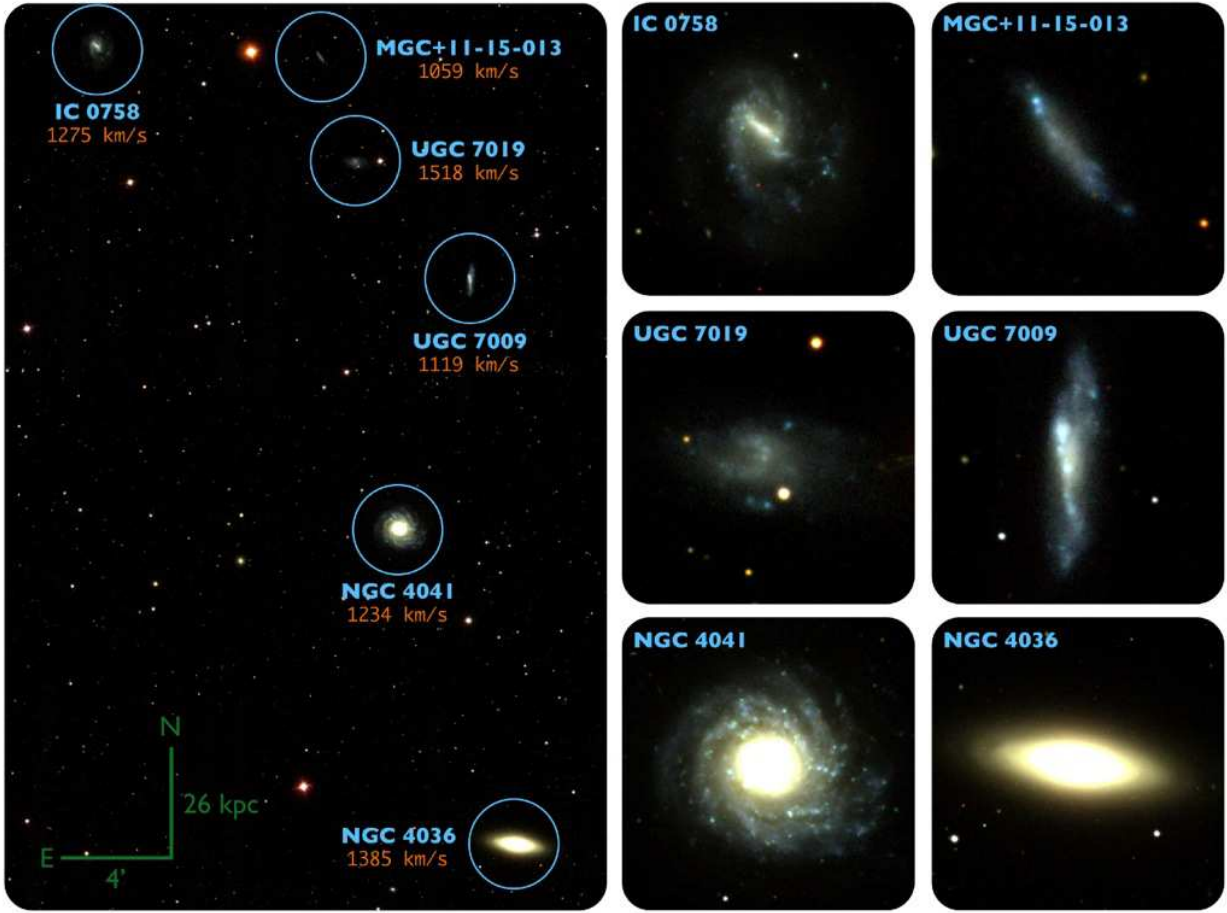}
		\caption{The loose grouping of galaxies around \n4041. 
		One is a distorted spiral, three appear irregular (albeit 
		the high inclination hinders this classification), and one 
		is an early-type galaxy with a faint inner dust lane 
		(evident in regular contrast scaling, but not here). We 
		mark the velocity of each galaxy on the SDSS mosaic image 
		on the left, and show high-contrast zoom-ins on the right 
		(not to scale). The four smaller galaxies (top two rows) 
		also exhibit considerable FUV fluxes on \gal\ images, 
		indicative of recent or ongoing star formation. Their 
		calculated SFRs are listed in Table~\ref{tab:sdss}. The 
		color of the inner disk of \n4041 appears similar to that 
		of the lenticular \n4036, while the outer disk shines 
		bright in shorter wavelengths. 
		}\label{fig:sdss}

	\end{center}
\end{figure*}

The complex structure of \n4041 could be the result of a past dynamical event, so we search the \textit{NASA Extragalactic Database} (\ned) for neighbors at accordant redshifts. We limit the search to galaxies in a cone of 1 degree about \n4041 and in a velocity space of $\Delta v < 400~$\kms. The search yields five neighbors, so the galaxy is by no means isolated. Details are given in Table~\ref{tab:sdss}, and a finding chart shown in Figure~\ref{fig:sdss}. 

% Figure: n4041 group
% ------------------------

The first five galaxies in this list comprise group LGG~266 in the listing of \citet{garcia93}. The sixth, MGC+11-15-013, also fits the LGG catalog limiting criteria -- a velocity difference of less than 600~\kms\ and a projected separation less than 0.52~Mpc -- but was only discovered recently \citep{mgc}. Since we find no morphological classification for this galaxy in the literature, we inspect its $z$~band SDSS image to find a linear profile with no pronounced central peak. We therefore classify it as Irr. The close-up of Figure~\ref{fig:sdss} shows a clumpy profile, but it is biased by bright star forming regions at the extremes of the disk that are not detectable in the $z$~band. 
The dominant galaxy of this group is \n4036, a lenticular. It features an inner equatorial dust lane and registers LINER emission \citep{veron06}.

% Table: n4041 group
% ------------------------
\begin{deluxetable}{lcccccr@{.}llc}
\tabletypesize{\small}
\tablewidth{0pt} 
\tablecolumns{7}
\tablecaption{The \n4036 Galaxy Group\label{tab:sdss}}
\tablehead{
\colhead{ID} & 					% C1
\colhead{RA} &					% C2
\colhead{DEC} &					% C3
\colhead{$m_g$} &				% C4
\colhead{$v_R$} &				% C5
\colhead{SFR} &					% C6
\tmult{$\log(M_*)$} & 	% C7, 8
\colhead{Morphology} &			% C9
\colhead{References}\\			% C10
& 												% C1
\colhead{(h~m~s)} & 								% C2
\colhead{($\circ$~$\prime$~$\prime\prime$)} &	% C3
\colhead{(mag)} &								% C4
\colhead{(\kms)} &								% C5
\colhead{(\Msun~yr$^{-1}$)} &					% C6
\tmult{(\Msun)} &										% C7, 8
& 												% C9
}
\startdata
\n4036        & 12~01~26.7 & +61~53~45 &  11.97 & 1385 & 0.02 & 10&42 & S0        & [1]~[1]~[2]\\
\n4041        & 12~02~12.2 & +62~08~14 &  12.19 & 1234 & 0.63 & 10&58 & SA(rs)bc? & [1]~[1]~[3]\\
IC~0758       & 12~04~11.9 & +62~30~19 &  14.25 & 1275 & 0.19 &  9&26 & SB(rs)cd? & [4]~[4]~[3]\\
UGC~7009      & 12~01~44.1 & +62~19~33 &  14.33 & 1119 & 0.12 &  8&78 & Im        & [4]~[4]~[4]\\
UGC~7019      & 12~02~29.4 & +62~25~02 &  15.05 & 1518 & 0.05 & \tmult{\nodata} &Im  & [5]~[6]~[7]\\
MGC+11-15-013 & 12~02~43.3 & +62~29~52 &  16.07 & 1059 & 0.03 & \tmult{\nodata} &Irr & [4]~[4]~[4]\\
\enddata
\tablecomments{Morphologies are taken from \citet{rc3}, apart from MGC+11-15-013, which we classify in Section~\ref{sec:group}. Stellar masses are derived though `best' 2MASS $K_S$ magnitudes \citep{2mass}, except \n4041, for which we use an aperture of $72\farcs0$ to match the measured aperture of the outer disk. SFRs are derived through \gal\ FUV imaging (see Section~\ref{sec:sfr}). SDSS photometry is model photometry in the $g$ band, where applicable. Positions, photometry, and radial velocities are drawn from the following sources, quoted in triplets in column 7: 
[1]~SDSS Data Release 6, \citet{sdss_dr6}; 
[2]~\citet{atlas3d_1}; 
[3]~\citet{rc3}; 
[4]~SDSS Data Release 2, \citet{sdss_dr2}; 
[5]~\citet{cotton99}; 
[6]~VATT $B$ band photometry, \citet{taylor05}; 
[7]~\citet{springob05}; 
The first five galaxies in this list make up group LGG~266 in the listing of \citet{garcia93}, and the sixth also obeys the inclusion criteria. Given the individual distance moduli, we derive absolute $g$~band magnitudes of $-20.0$, $-19.6$, $-17.9$, $-17.3$, $-16.7$, and $-15.0~$mag. 
}
\end{deluxetable}\vspace{-5pt}

% Subsection: n4041 fork
% -----------------------------------------
\subsection{Extended Structure}\label{sec:fork}
\begin{figure}[!t]
	\begin{center}

		\includegraphics[width=0.46\textwidth]{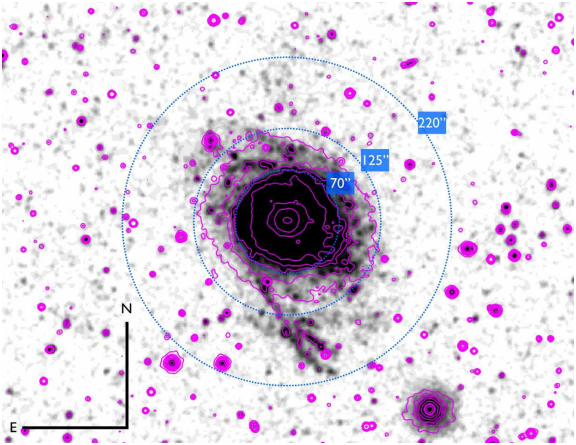}
		\caption{
			FUV+NUV image from \gal, smoothed with a three-pixel 
			Gaussian kernel, and presented at high contrast to 
			emphasize low surface brightness features. We reveal 
			the presence of material beyond even the extended UV 
			disk of \n4041,  in the form of a three-pronged 
			structure. This `tidal fork' might be part of a 
			greater network of debris, but it is the only feature 
			detected at a level of 3-5~$\sigma$ above the 
			background. The contours show the isophotal structure 
			of a $g+r$ image from SDSS at the 1, 2, 3, 5, 10 
			$\sigma$ level. Only the brightest, central prong is 
			detectable in SDSS at the 1-2~$\sigma$ level. The 
			absence of H$\alpha$ emission (transmitted in the 
			$r$~band filter) suggests an age on the order of 
			100~Myr. Concentric circles mark the outer disk of 
			\n4041, a faint, extended UV disk, and the full radial 
			extent of the fork. The angular distances correspond 
			to 9, 14, and 24~kpc at the distance of 22.7~Mpc. 
			The bright feature in the southwest is a foreground 
			F-type star. 
		}\label{fig:fork}

	\end{center}
\end{figure}

Upon inspecting \gal\ UV imaging of \n4041 we find a faint, three-pronged feature to the south of the UV disk. It extends between $125^{\prime\prime}$ and $220^{\prime\prime}$, or $\sim14-24~$kpc. This `tidal fork' is shown in the far-ultraviolet plus near-ultraviolet (FUV+NUV) image of Figure~\ref{fig:fork}. We stack the two \gal\ channels to boost the flux, and smooth by a three-pixel Gaussian kernel to remove noise. We present this image at high contrast to accentuate low surface brightness features. The contours mark optical brightness, derived by stacking the $g$ and $r$ SDSS images, and smoothing by a ten-pixel Gaussian kernel.

% Figure: tidal fork
% ------------------------

The feature is detectable in the UV image at the 3-5~$\sigma$ level, but not quite as bright in the stacked and smoothed optical image ($\lesssim2~\sigma$ detection). Assuming this is a tidal feature physically associated with \n4041, we can attribute it to a past interaction. The low $g+r$ flux denotes low or no H$\alpha$ emission, and therefore no current star formation in the `fork'. The UV flux should then mostly originate from activity on the order of 100~Myr ago, rather than current star formation and O-stars.

% Subsection: n4041 (s)SFR
% -----------------------------------------
\subsection{Star Formation Rate, Stellar and Gas Mass}\label{sec:sfr}
Since there are no published values for the star formation rate of \n4041, we apply the \citet{kewley02sfr} methodology to obtain estimates from the \textit{IRAS} infrared fluxes at $60~\mu$m and $100~\mu$m. In brief, we first employ the \citet{helou88} prescription to obtain the far-infrared flux, $F_\fir$, from the \textit{IRAS} fluxes, which we then convert to a luminosity through $L_\fir = 4\,\pi\,{D}^2\times F_\fir$, where $D$ is the distance to the galaxy. We then use the \citeauthor{kewley02sfr} adaptation of the \citet{kennicutt98} law to derive a FIR star formation rate (SFR) of 4.10~\Msun~yr$^{-1}$ for \n4041. This compares well with galaxies of similar morphological type and brightness in the SINGS survey \citep{sings}. The flux ratio between 60 and 100~$\mu$m of 0.44 is also in accord with SINGS. 

We also obtained an FUV SFR for \n4041 through archival \gal\ imaging, which we expect to be lower, due to the absorption of UV photons by dust. To estimate the flux from the target we fit it with the \texttt{ELLIPSE} function in IRAF and added up the counts corresponding to the two segments of the galaxy disk (assuming the above radii of $80^{\prime\prime},~22^{\prime\prime}$ and no ellipticity), while accounting for a flat background level and a foreground extinction of $E(B-V)=0.15~$mag \citep[following $A_\textup{\scriptsize FUV} = 7.9\times E(B-V)$ from][]{gildepaz07}. Note that the low resolution of \textit{IRAS} images does not allow for a similar treatment. We then convert fluxes to star formation rates of 0.18,~0.65~\Msun~yr$^{-1}$ through the FUV \citet{kennicutt98} relation. Considering the area of each segment, we derive SFR densities of 
$[0.45,~0.08]~$\Msun~yr$^{-1}$~arcmin$^{-2}$, or 
$[1.03,~0.20]\times10^{-2}$~\Msun~yr$^{-1}$~kpc$^{-2}$
for the inner disk and outer galaxy respectively. We apply the above methodology to also derive the FUV SFRs of the group around \n4041, and list results in Table~\ref{tab:sdss}.
The observation of a markedly higher SFR density in the central regions of a galaxy is common in post-interaction systems, where gas is funneled inwards \citep{forster03,isk09a} until it is the nucleus alone that is experiencing a starburst, or until it ignites an active nucleus \citep{ellison11}.

Stellar masses, $M_*$ were derived by applying the \citet{bell03} prescription with a $K_S$ solar brightness of 3.32~mag to catalogued 2MASS $K_S$ photometry \citep{2mass}. The faintest two members of the group are below the detection limit, and the obtained values for the four brighter galaxies are listed in Table~\ref{tab:sdss}. While we generally use the 2MASS `standard' radius, in the case of \n4041 we use a radius of $72^{\prime\prime}$ in order to envelop the entirety of the outer galaxy. This gives rise to a larger $M_*$ for \n4041 than \n4036, despite the brighter tabulated $g$ band magnitude of the latter. Additionally, we derive a specific SFR (sSFR) of \n4041, through division with the FIR SFR, of $10.9\times10^{-11}~\textup{yr}^{-1}$, consistent with the morphological type of \n4041 -- \cf\ its near morphological counterparts HCG~7C and HCG~59A, that also register values $\simeq10^{-10}~\textup{yr}^{-1}$ \citep{isk10,isk12a}. 

Finally, we look for information on the gaseous component of \n4041. The survey of \citet{couto89} offers a value of $\log (M_\textup{\scriptsize H\one}/\Msun) = 9.76$. We also use the CO luminosity published by \citet{elfhag96} to estimate the mass of molecular hydrogen in the system. We use the classical Milky Way conversion factor, $\alpha_\textup{\scriptsize CO}=4.6~$\Msun\ \citep{solomon87}, to convert the $\log (L_\textup{\scriptsize mol})=8.48~$K~\kms~pc$^2$ to a gas mass of $\log (M_\textup{\scriptsize mol}/\Msun) = 9.14$. The combined gas mass, $M_\textup{\scriptsize gas} = 1.3~\mhimath+M_\textup{\scriptsize mol}$ \citep[the factor of $0.3~\mhimath$ accounts for helium, \eg][]{hunter82}, is therefore $\log (M_\textup{\scriptsize gas}/\Msun) = 9.95$. Divided by the SFR, we get a gas depletion timescale of $\sim2.2~$Gyr.  

% Subsection: Interaction? 
% -----------------------------------------
\subsection{A Possible Dynamical Event in the Recent Past}\label{sec:dyn}
Many of the traits examined in this section are consistent with a past dynamical event in \n4041: a segmented brightness profile; a double central peak, offset from the isophotal center; a UV-bright tidal feature; and a markedly higher central SFR. Of particular interest is the offset central peak, which potentially represents the dynamically decoupled core of \n4041 \citep[following][]{marconi03}. Such features herald past mergers \citep[\eg][]{Emsellem11}, not interactions. 

The group environment is thought to be conducive to dynamical events and accelerated galaxy evolution \citep[\eg][]{vm01,johnson07,gallagher10,isk10,isk12a}. Furthermore, while a dense environment can enhance star formation during interactions \citep{martig08}, the effects of minor events are often not very pronounced, leaving but the faintest traces of their passing \citep{isk10}. At the same time, compact galaxy groups have been nominated as the sites of lenticular galaxy formation \citep{wilman09}, as minor mergers build up mass and gradually exhaust the gas reservoir of the group, while retaining the structure of individual disks. The high-index S\'ersic outer brightness profile for \n4041 (Figure~\ref{fig:sbp}) is more akin to a lenticular than a spiral. 

The overall symmetry of the disk seems to rule out a major merger. We therefore propose an accretion event, perhaps that of a gas-rich dwarf, as the origin of the present star formation activity and the two-component disk of \n4041.

%======= 5. STAR CLUSTER ANALYSIS =======
\section{The Bimodal Star Cluster Population of \n4041}\label{sec:clusters}
In the Introduction of Section~\ref{sec:intro} we argued that the understanding of a galaxy and its surroundings can be enhanced by a study of the star cluster population. We now present such a study for \n4041. We split the sources according to galactocentric radius (\rgc), in order to contrast the sub-populations of the inner and outer regions, and derive analysis only from high-confidence detections, \ie\ sources flagged in Section~\ref{sec:data-products} as \fubvi~$=1111$. 

%%% SUBSECTION ----
\subsection{Star Cluster Colours}\label{sec:sc:colours}
\begin{figure}[!h]
	\begin{center}
		\includegraphics[width=0.47\textwidth]{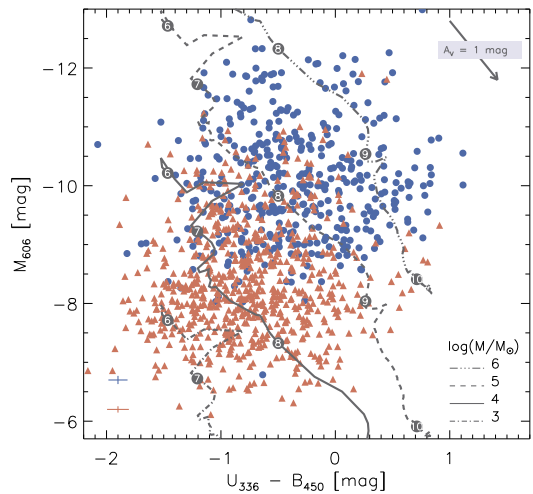}
		\caption{
		Color-magnitude diagrams of high-confidence cluster 
		candidates with \ygg\ models of solar metallicity 
		overplotted on isotropic axes. Model tracks represent 
		masses of $10^3$, $10^4$, $10^5$, and $10^6~$\Msun\ 
		from bottom to top, each tracing ages between 6~Myr 
		and $\sim10~$Gyr. Numbers in filled circles mark each 
		age dex of model SSP evolution. There is a pronounced 
		difference in the color distributions of clusters in 
		the inner disk (blue dots) and outer galaxy 
		(red triangles), likely a combination of differing 
		detection limits and possibly also mass distributions. 
		Such disparities are often observed in interacting 
		systems, but not elsewhere. The crosshairs in
		the bottom left represent the median error in each 
		axis. 
		}\label{fig:cmd}
	\end{center}
\end{figure}

The colors of star clusters are indicative of the age of the stellar population they represent, hence SSP models can be used to interpret the colors of cluster candidates across \n4041. The  color-magnitude diagram (CMD) of Figure~\ref{fig:cmd} plots the evolution of SSP models of various masses (from $10^3$ to $10^6$~\Msun), and shows dissimilar distributions for clusters in the inner and outer galaxy (blue circles/red triangles). While this hints at different mass functions, incompleteness will eliminate the low-luminosity end of the inner disk distribution. Still, such dramatic contrasts between subpopulations within a single system are not normally observed outside merging and interacting galaxies \citep[\eg][]{fedotov11}. We remind that no spatial trends were discovered in the metallicity distributions of clusters in the inner and outer galaxy (see Section~\ref{sec:sed}), and note that the reddening distribution shows no change between inner disk and outer galaxy.

% Figure: CMD
% ------------------------
% ------------------------

\begin{figure*}[!t]
	\begin{center}

		\includegraphics[width=\textwidth]{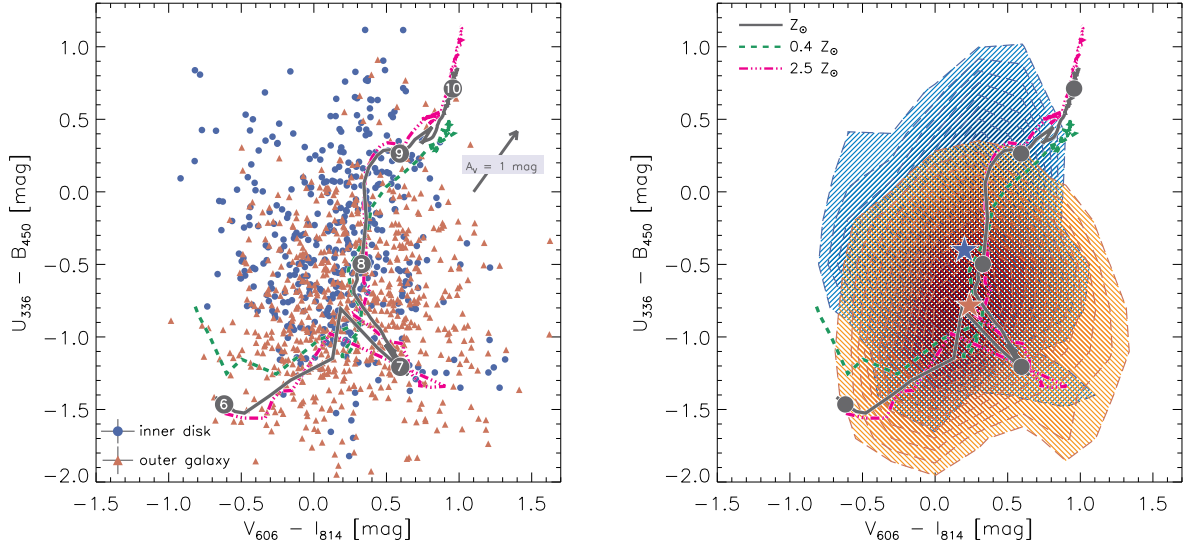}
		\caption{
		{\bf Left:}
		$U_{336}-B_{450}~vs~V_{606}-I_{814}$ colors of star 
		cluster candidates in \n4041 on isotropic axes. We 
		overplot \ygg\ models, following the conventions of 
		Figure~\ref{fig:cmd}, and also include sub- and 
		super-solar tracks in this case. Clusters in the 
		inner disk and outer galaxy are marked as blue circles 
		and red triangles respectively, while the crosses 
		on the lower left denote the median photometric error 
		of each subsample. The extinction vector marks one 
		magnitude of attenuation in the V band \citep{cardelli89}. 
		Sources in the inner disk often diverge from the 
		model track toward bluer values (top left), a likely 
		effect of crowding. This could alternatively be 
		attributed to differing filling factors in the gas, 
		in which the emission originates (see 
		Section~\ref{sec:sed}).
		\mbox{{\bf Right:}
		Two-dimensional} color-density plot for the inner disk 
		(blue), and outer galaxy (red). We omit areas that 
		sample less than 15\% of the 2D-histogram peak. The two 
		distributions appear slightly offset, extending each 
		other in opposite directions. The median colors 
		(marked as stars) are separated by $\approx0.5~$mag, 
		possibly indicating a different age distribution for 
		the two sub-populations in recent times (completeness 
		affects the old end of the model track). 
		}\label{fig:colours}

	\end{center}
\end{figure*}

We further investigate this disparity through the color-color plot of Figure~\ref{fig:colours}. As in the CMD of Figure~\ref{fig:cmd}, the model track spans the full evolution of an SSP, from 1~Myr to about 10~Gyr. The SSPs account for the transmission of nebular continuum and emission lines in the F606W filter. This extends the purely stellar SSP toward greener values and therefore cover sources that scatter toward the top left of the color distribution. We show color contours on the right panel to aid comparison, where blue and red shades represent the inner/outer galaxy. To avoid over-representing outliers we omit the lowest two contour levels, hence plotting only two-dimensional bins that sample at least 15\% of the full population. Sources in the inner disk appear to often diverge toward bluer values. We attribute this to crowding and aperture effects: on the vertical axis, the \bb\ flux is measured from WFPC2, at lower resolution; and on the horizontal axis, we might expect some contamination by the diffuse, irregularly distributed \ha\ emission in the inner disk. In addition, the colors of low-mass clusters might preferentially diverge toward this region of colorspace \citep{fouesneau10,popescu10,silvavilla11}. A third possibility is a variety of filling factors in the gas, which might lead to variations of up to 0.5~mag in the model tracks. Since we adopt a given filling factor and electron density, we might be underestimating the blue emission from these regions.

The loci of the color contours (on the right), indicated by stars, are offset by $\approx0.5~$mag, while the overall distributions extend each other: younger ages are under-represented in the inner disk, and the outer galaxy shows few clusters older than 1~Gyr. This appearance is likely due to selection effects, namely the different detection limits in the inner disk and outer galaxy. On the young end, crowding in the inner disk might be responsible for this apparent lack of sources, especially ones with low mass. Furthermore, the higher SFR of the inner disk (Section~\ref{sec:sfr}) is expected to correlate with the number of young clusters. 

Such a disjointed color distribution might be expected in a merger, \cf\ the cluster populations of the Antenn\ae~\citep{whitmore95}, \n3256~\citep{trancho07b,trancho07a}, and most notably \n7252, which exhibits an inner star forming disk and an outer halo-type distribution~\citep{miller97,schweizer98}. Even so, such an orderly inner-versus-outer galaxy distinction has not been noted in past analyses. 
%

% Figure: CC
% ------------------------
% ------------------------

%%% SUBSECTION ----
\subsection{Luminosity Function}\label{sec:sc:LF}
\begin{figure*}[!p]
	\begin{center}

		\includegraphics[width=0.32\textwidth]{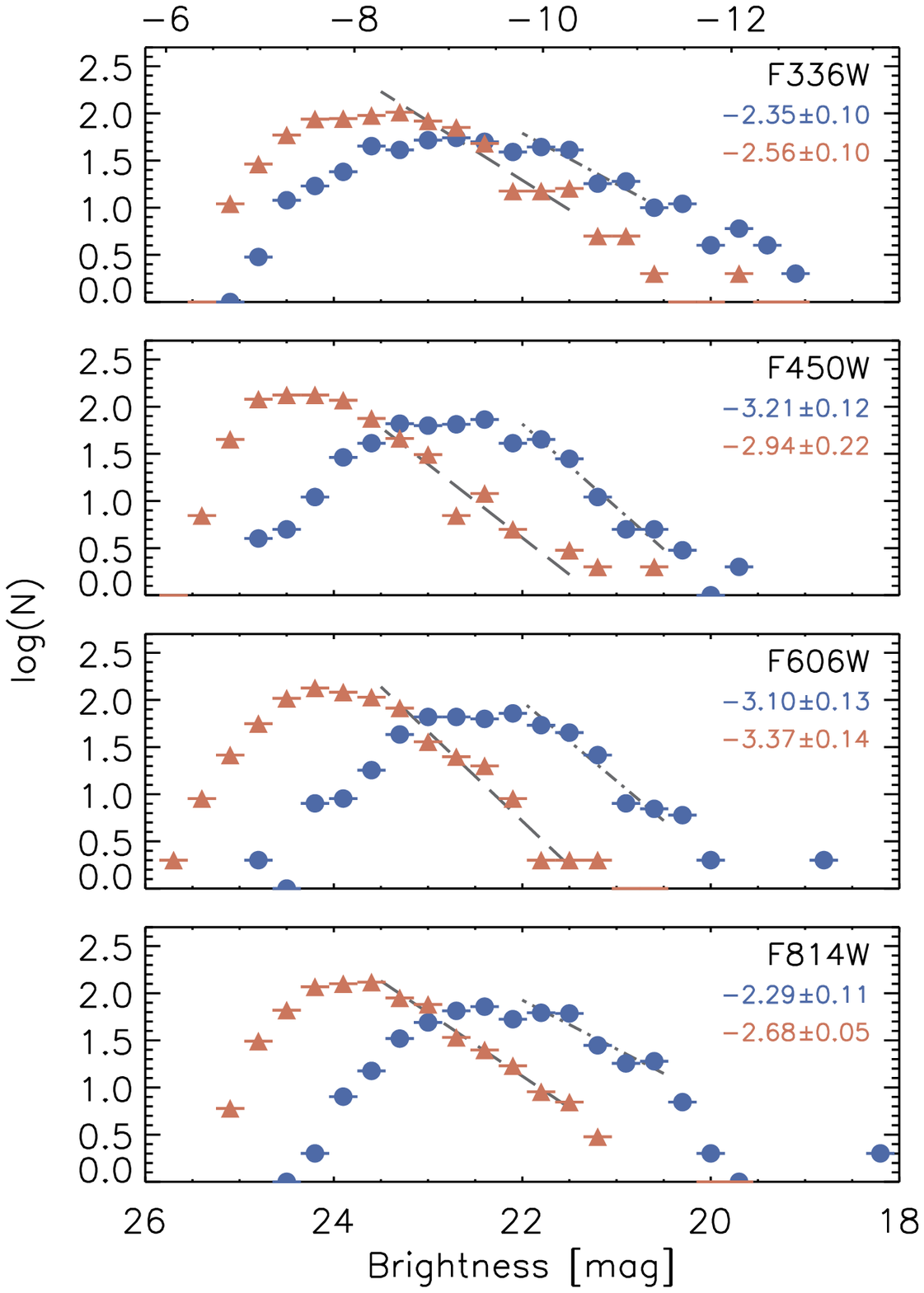}
		\includegraphics[width=0.31\textwidth]{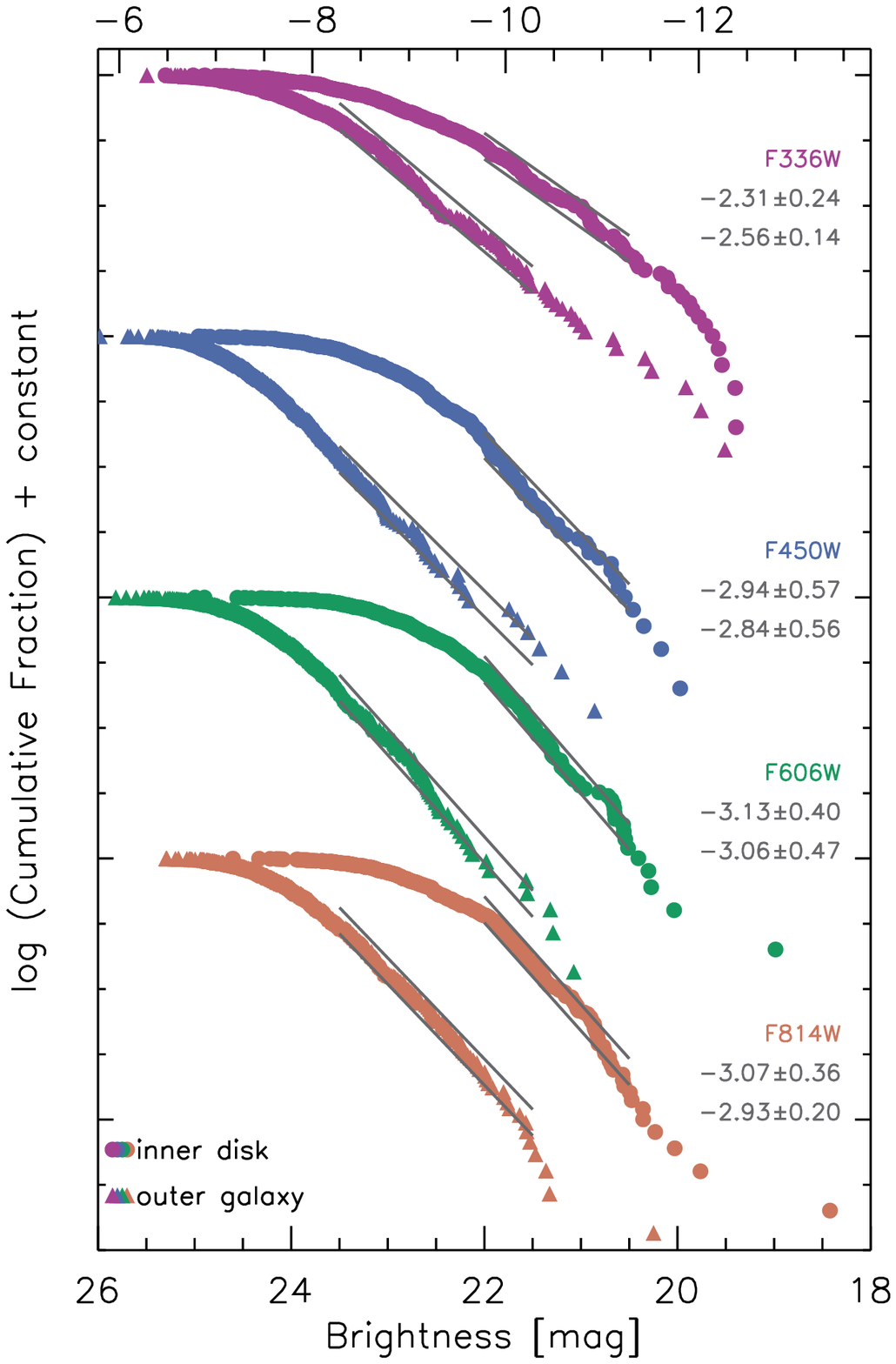}
		\includegraphics[width=0.31\textwidth]{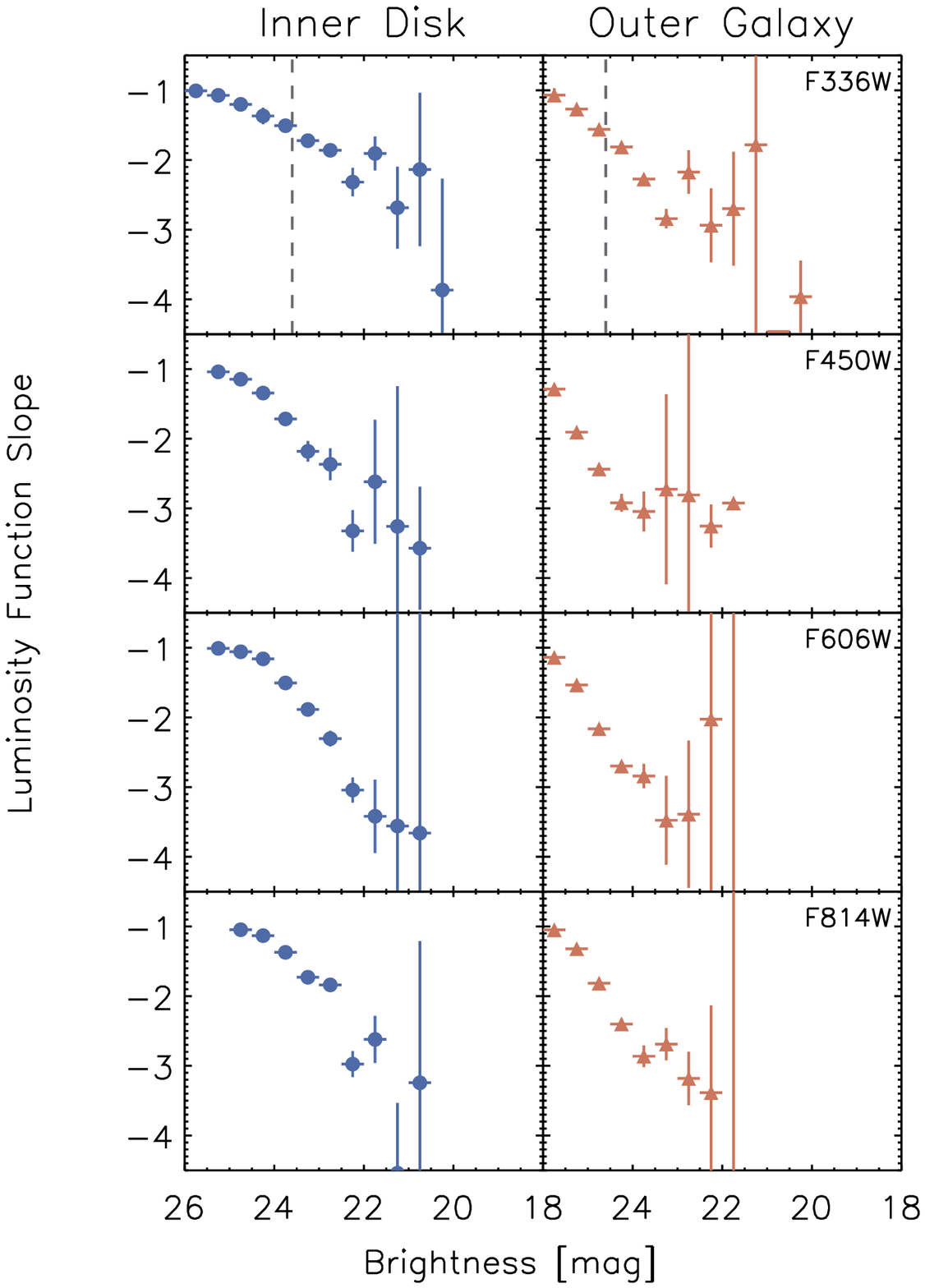}
		\caption{Binned ({\bf left}) and cumulative ({\bf right}) 
			luminosity functions (LF) of star cluster candidates 
			in \n4041, arranged according to filter and \rgc\ 
			(inner disk, outer galaxy). The lines (bands on the 
			right) denote the range over which we fit the slope 
			of each function, and we note slopes next to each pair 
			of LFs. The \uub\ LF of the inner disk is shallower 
			than other functions, which are generally consistent 
			within the errors. A higher overall extinction would 
			not explain this discrepancy. Slopes steeper than $-2$ 
			are to be expected, given the bright absolute magnitudes 
			being fit \citep[][see text for details]{gieles10}. 
			Binning smears the substructure evident in the 
			cumulative functions and is subject to sampling errors. 
			{\bf Right}: LF slopes, measured on the cumulative 
			functions of the center panel. Dashed lines indicate 
			the 90\% 
			completeness limits. As expected by the literature 
			review of \citet{gieles10}, the slope becomes steeper 
			with increasing brightness. We attribute this to the 
			segmented structure of the function, as it gives 
			rise to shallow slopes in the incomplete part, 
			values consistent with the literature in the 
			power-law segment, and very steep slopes once 
			the upper truncation sets in. The increasing errors
			reflect the sampling (decreasing numbers). 
		}\label{fig:LF}

	\end{center}
\end{figure*}

The Luminosity Function (LF) of star cluster populations is routinely observed to have a smooth, power-law shape, as noted in the Introduction. While this shape can be seen in the binned LF of Figure \ref{fig:LF} (left), the cumulative functions on the right panel reveal a striking substructure of: (i)~an incomplete part with an evolving gradient; (ii)~a smooth, power-law segment; and (iii)~a truncation at the bright end. This indicates that substructure may have been habitually ignored in the literature by arranging clusters in bins of arbitrary width \citep[see the cautionary note of][]{jesus05}. 

% Figure: LF
% ------------------------
%
%%% LF TABLE
\input{tab_LF.tex}

We measure the slope from the cumulative function, in two ways. First we fit only the smooth part of each function, but maintain the fit consistent between the four bands. Then we fit over a large range, in order to emulate the effect of binning. The results, listed in Table~\ref{tab:LF}, show shallower functions for the Ôfull rangeÕ fit (consistent with the literature but slightly steeper than the canonical value of $-2$), while selective fits of the power-law segment return steeper values. Fitting only the power-law part of each function provides the added advantage of not biasing the measurement according to physical parameters such as cluster age or reddening. Since young clusters tend to be much brighter than older ones, the bright end of the LF will largely represent the young part of the population. Our tests indicate that fitting the entire range can lead to very large errors and possible biases that are not present when fitting only the power-law segment. 

The \uub\ returns markedly shallower fits than the rest of the bands, an effect noted in the literature review of \citet{gieles10}, where redder bands were shown to have steeper distributions. We note that the extinction does not change the slope of the LF \citep{larsen02lf}. The \citeauthor{gieles10} analysis also showed LF slopes to be a function of the mean luminosity. The mean brightness over which we fit (21 or 22~mag; see Table~\ref{tab:LF}) translates to a luminosity of $\log(L/\Lsun)\simeq5$, corresponding to literature slopes of $-2.5$. We attribute the consistently steeper slopes derived from the binned histograms to sampling uncertainties, especially given the relatively small cluster population being fit. This discrepancy exemplifies the loss of information that occurs when binning LFs, while confirming the \citeauthor{gieles10} relation: the farther `up' the function we move, the steeper the values become. 

To confirm this link to the \citeauthor{gieles10} we plot the slope of the LF as a function of brightness. We measure in 0.5~mag bins across the range shown in the LFs of Figure~\ref{fig:LF} and find the slope to become steeper in proportion to the brightness. This evolution of the LF slope is shown in Figure~\ref{fig:LF}. Our results therefore suggest that all LFs might display substructure \citep[also seen in the cumulative LFs of][]{bastian12}, which would account for the \citeauthor{gieles10} relation, especially given its correlation with the range being fit. 

The LFs also provide a test of photometric completeness. As outlined in Section~\ref{sec:data-products}, we select in F336W but apply filters that expect the detection of sources across the \ubvi\ baseline. This implicitly includes the \bvi\ bands in defining the completeness of our sample. In order to estimate this effect, we plotted the LFs by omitting the criteria pertaining to multi-band photometry and found the shape of each function to be unchanged by this process. The measured slopes were consistent within the errors with those listed in Table~\ref{tab:LF}. We also examined the effect of potentially age-dependent reddening, by repeating the above exercise after limiting the age and extinction of the sample being fit. In all cases, the derived slopes were consistent with the fits of Table~\ref{tab:LF}.

%% Figure: LF slope
%% ------------------------
%\begin{figure}[tbhp]
%	\begin{center}
%
%		\includegraphics[width=0.47\textwidth]{f15.eps}
%		\caption{The slope of the LFs of Figure~\ref{fig:LF}. 
%			Dashed lines indicate the 90\% completeness limits. 
%			As expected by the literature review of 
%			\citet{gieles10}, the slope becomes steeper with 
%			increasing brightness. We attribute this to the 
%			segmented structure of the function, as it gives 
%			rise to shallow slopes in the incomplete part, 
%			values consistent with the literature in the 
%			power-law segment, and very steep slopes once 
%			the upper truncation sets in. The increasing errors
%			reflect the sampling (decreasing numbers). 
%		}\label{fig:LFslope}
%
%	\end{center}
%\end{figure}

%%% SUBSECTION ----
\subsection{Mass Function}\label{sec:sc:MF}
\begin{figure}[!h]
	\begin{center}

		\includegraphics[width=0.47\textwidth]{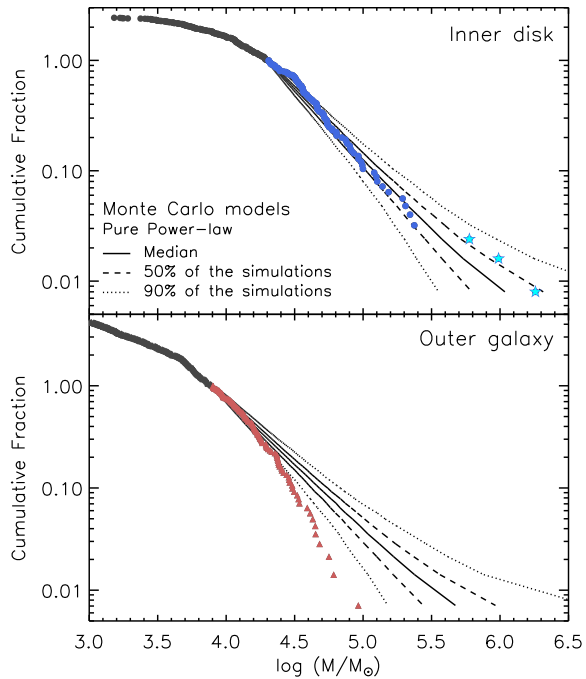}
		\caption{Mass functions (MFs) of star clusters in the 
			inner disk and outer regions of \n4041. The MFs cover 
			only clusters aged $10-100~$Myr and thus avoid fitting 
			on young, unbound associations, and incomplete samples 
			at older ages. Both MFs show a composite structure in 
			three parts: an incomplete low-mass section, followed by 
			a power-law, which is truncated above 
			$\log(M_*/\Msun)\approx4.4$ in the outer galaxy. The 
			lines represent a range of modeled pure power-law MFs 
			of exponent $-2.3$ (see text), which do not provide an 
			adequate description of the observed function for the 
			outer galaxy, but describe the inner disk well. The 
			light blue stars represent the three nuclear clusters 
			in \n4041, as discussed in Section~\ref{sec:n4041}.
		}\label{fig:MF}

	\end{center}
\end{figure}

In order to quantify the cluster mass distributions in the inner and outer disk, we plot the Mass Function (MF) in Figure~\ref{fig:MF} as a set of cumulative distributions (counting from high-to-low mass). This plot will suffer from various incompleteness effects at different star cluster ages. On the young end we need to exclude short-lived, unbound stellar associations that masquerade as star clusters at large distances \citep{gieles11a}. Toward older ages, the diagram is affected by the evolutionary fading and reddening of SSPs. This decreases the mass-to-light ratio, therefore increasing the mass required to detect a star cluster with increasing age. These evolving detection limits are represented by the dotted and dashed lines (inner/outer galaxy) on Figure~\ref{fig:MF}, and shape this parameter space. 

% Figure: MF
% ------------------------
% ------------------------

We therefore restrict the MF to cluster candidates aged between [10, 100]~Myr. We note an offset of $\simeq 0.5~$dex between the inner disk and outer galaxy, however the two functions display consistent slopes. Interestingly, the double-break structure reported for the LFs (Figure~\ref{fig:LF}) persists in the MF, which displays an incomplete part, a power-law segment, and an upper truncation. 

We investigate the origin of this segmentation by comparing to a set of power-law models, represented by the various markings on Figure~\ref{fig:tM}. The slope is deduced as $\alpha=-2.3$ through this process, consistent with the literature-standard index of $-2.0$. In brief, we follow the methodology of \citet{bastian12} and run a series of models of the MF with a sample size equal to the observed sub-populations, and assume a power-law distribution. The slope is derived after few trial-and-error iterations. The median of all models is shown as a solid line, and dashed and dotted lines indicate the range containing 50\% and 90\% of all modeled outcomes. The observed functions are seen to diverge from the power-law models at the low-mass end, {]bf and also the and high-mass end for the outer galaxy}. While incompleteness will affect low-mass clusters, at the high-mass end we expect to detect the vast majority of sources. This divergence is therefore established as a physical effect and the MF of the outer galaxy interpreted as a Schechter-type distribution with a truncation at $\log(M_*/\Msun)\approx 4.4$. This follows on the analyses by \citet{gieles09a} and \citet{larsen09} that favored Schechter MFs, and the truncation mass is in accord with the value derived for M83 by \citet{bastian12}. As with the LF, the three-part structure of the MF suggests that binning acts to erase the information imprinted in the mass distribution. 

Interestingly, the MF of the inner disk is consistent with the power-law model, an effect potentially related to the higher SFR in that region. The light blue stars of Figure~\ref{fig:MF}, top, represent the three nuclear region clusters discussed in Section~\ref{sec:n4041}.

%%% SUBSECTION ----
\subsection{Age Distribution and Cluster Disruption}\label{sec:sc:ages}
\begin{figure}[!h]
	\begin{center}
		\includegraphics[width=0.53\textwidth]{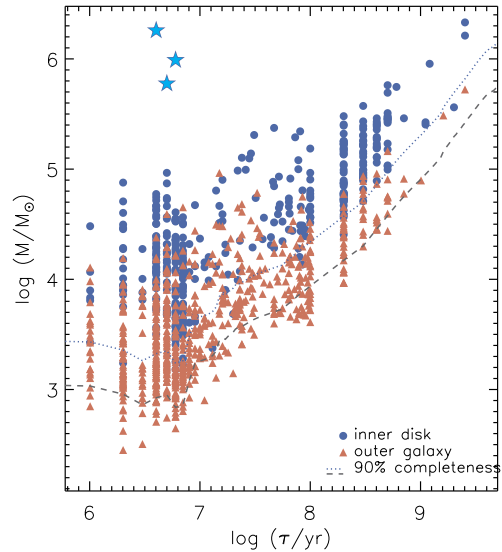}
		\caption{Age-mass diagram for cluster candidates in the 
			inner disk (blue circles) and outer regions (red 
			triangles) of \n4041. 
			The three young, massive sources in the inner disk 
			(cyan stars) are located in the nuclear region, and 
			may not follow the same scaling laws as SSPs. The 
			lines show the 90\% completeness limit translated 
			into the mass of an \ygg\ model across the given 
			range of ages. This limit evolves according to the 
			temporally increasing mass-to-light ratio of SSPs. 
			The two sub-populations are offset by approximately 
			0.5~dex in mass, which follows on the relative shift 
			of Figure~\ref{fig:MF}. The age distributions appear 
			consistent and typical of their late-type host galaxy. 
			}\label{fig:tM}

	\end{center}
\end{figure}

Combining mass with age we obtain the diagram of Figure~\ref{fig:tM}. The relative shift in mass noted above (Section~\ref{sec:sc:MF}) is obvious here as a vertical shift according to different detection limits, while the age distributions appear largely similar and not unlike others studied in the past. The three inner disk outliers at high mass and young age represent the double nuclear peak and another, nearby source. Assuming these are nuclear star clusters, we do not expected them to follow normal scaling laws \citep{seth08,scott12}; as mixed stellar populations their mass-to-light ratios will be unlike those of SSPs. 

From the first age dex, $\log(\tau/yr)\in[6, 7]$, we can make a rough assessment of the state of star formation in \n4041. While the outer galaxy might have recently produced a larger overall number of clusters above the detection limit, the inner disk is consistently producing more massive clusters (the detection limit there is also brighter). The correlation between the mass of the most massive cluster in a system and the SFR \citep{bastian08sfr} indicates that the rate is higher in the inner disk, in accord with the value derived from the FUV fluxes (as noted in Section~\ref{sec:sfr}). 

% Figure: age-mass
% ------------------------
% ------------------------

Interpreting the distribution at older ages is more complicated. The formation of a star cluster does not guarantee its perpetual existence, as star clusters face many disrupting processes over their early evolution \citep[see][for an overview]{bastian-gieles08}. In that way the presence of a cluster in a certain age bin is modulated not only by the fading lines discussed above, but also its survival as a physical entity. To decouple evolutionary fading from disruption, we can fit the age-mass distribution with a suite of models of mass-dependent cluster disruption. A dependence on mass and environment was favored by the fits of \citet{bastian12} and we follow this methodology here. Other flavors of cluster dissolution -- \eg\ mass-independent \citep{fall05}, or purely environment-dependent \citep{elmegreen10} -- or combinations thereof, may also be valid. 

\begin{figure*}[!t]
	\begin{center}

		\includegraphics[width=0.8\textwidth]{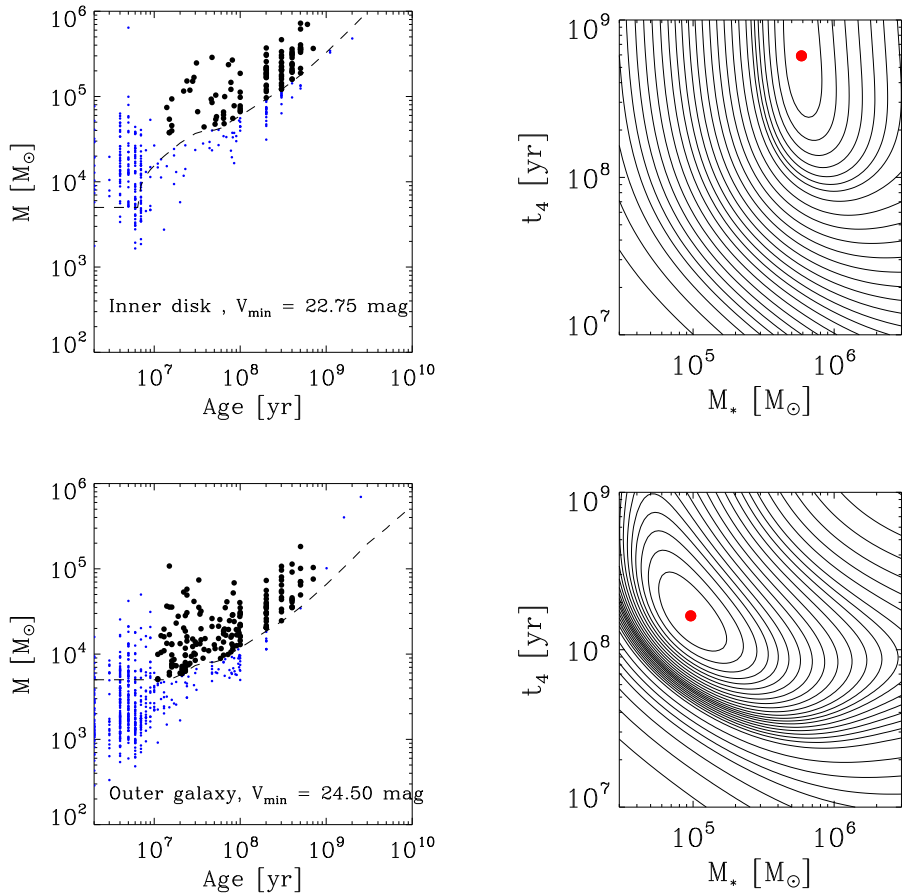}
		\caption{Disruption time for a $10^4~$\Msun\ cluster 
			($t_4$) versus Schechter Mass ($M_*$, where the 
			truncation sets in) for the inner disk and outer 
			regions of \n4041. This plot uses the age-mass 
			diagram of Figure~\ref{fig:tM} and completeness
			(based on an \ygg\ SSP model) to estimate the two 
			quantities through a series of maximum-likelihood 
			tests. Similarly to the MFs of Figure~\ref{fig:MF}, 
			only the black dots are used, in order to avoid 
			incompleteness effects and short-lived stellar 
			associations. We find values consistent with the 
			literature in the outer galaxy, while the atypically 
			high $M_*$ derived for the inner disk might be 
			affected by crowding and incompleteness. This is 
			evident from the shallow gradient of the contours, 
			\ie\ the large range of values that fit the data 
			well. 
		}\label{fig:t4}

	\end{center}
\end{figure*}

We perform a maximum likelihood analysis of the age and mass distributions, shown in Figure~\ref{fig:t4}, given the cluster disruption formulations of \citet{lamers05} and \citet{gieles09a}. This way we simultaneously compute $M_*$ and $t_4$, the dissolution time of a $10^4~$\Msun\ cluster. We employ a Schechter function and the modeled evolution of an \ygg\ SSP in \vb, the limiting filter (see Section~\ref{sec:completeness}, Figure~\ref{fig:err4}). As with the LF analysis, we exclude sources younger than 10~Myr to avoid stellar associations, and take into account incompleteness at older ages (through the \ygg\ model). This leaves the black dots of Figure~\ref{fig:t4}, which are fit with models based on a range of variations of $M_*$ and $t_4$.

% Figure: t4 vs Mstar
% ------------------------
% ------------------------

The fits favor a $t_4$ of a few hundred Myr in the outer galaxy and a longer time in the inner disk, contrary to \citet{lamers05}, who found environment density to correlate inversely with $t_4$. However, the small number of data-points in the inner disk provides broad confidence contours, expressed as shallow topography, therefore no deductions should be made from this fit. The $M_*$ seems to change significantly with local environment (inner/outer galaxy), in a manner consistent with the M83 study from which this methodology is derived: denser regions display higher $M_*$.

%%% SUBSECTION ----
\subsection{Star Formation History}\label{sec:sc:dndt}

\begin{figure*}[!t]
	%\begin{center}
		%
		\includegraphics[width=0.49\textwidth]{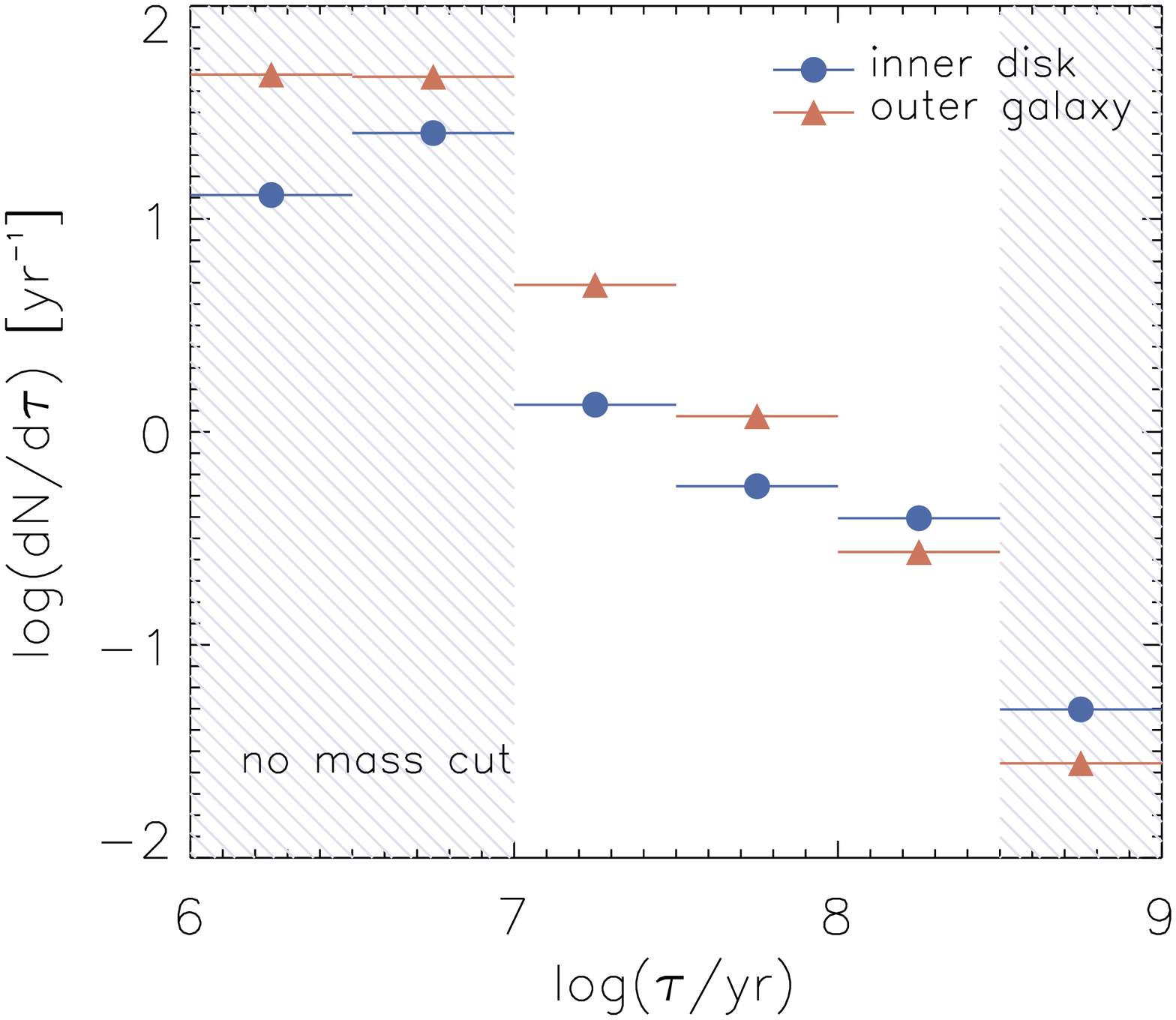}
		\includegraphics[width=0.49\textwidth]{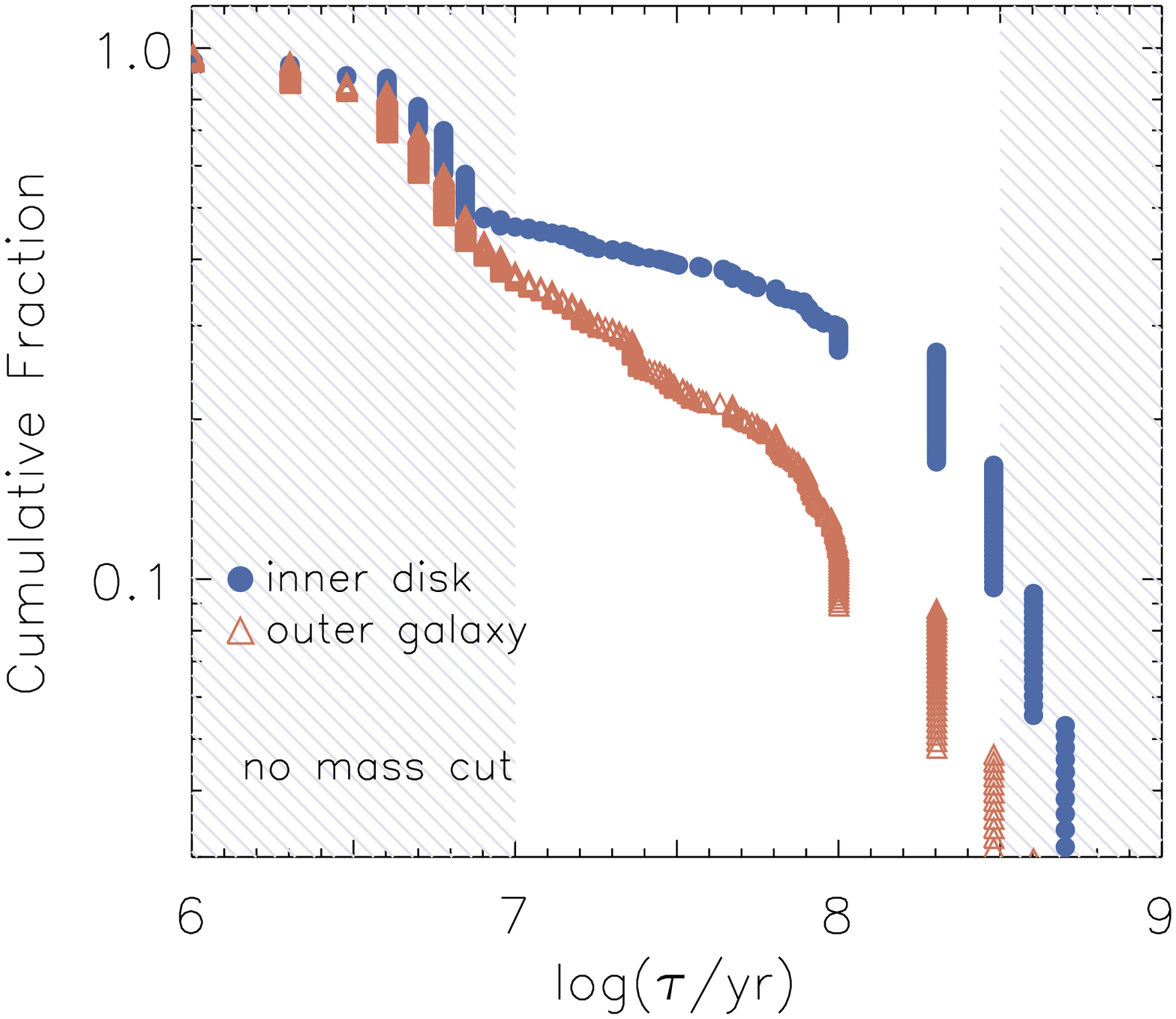}\\
		\includegraphics[width=0.49\textwidth]{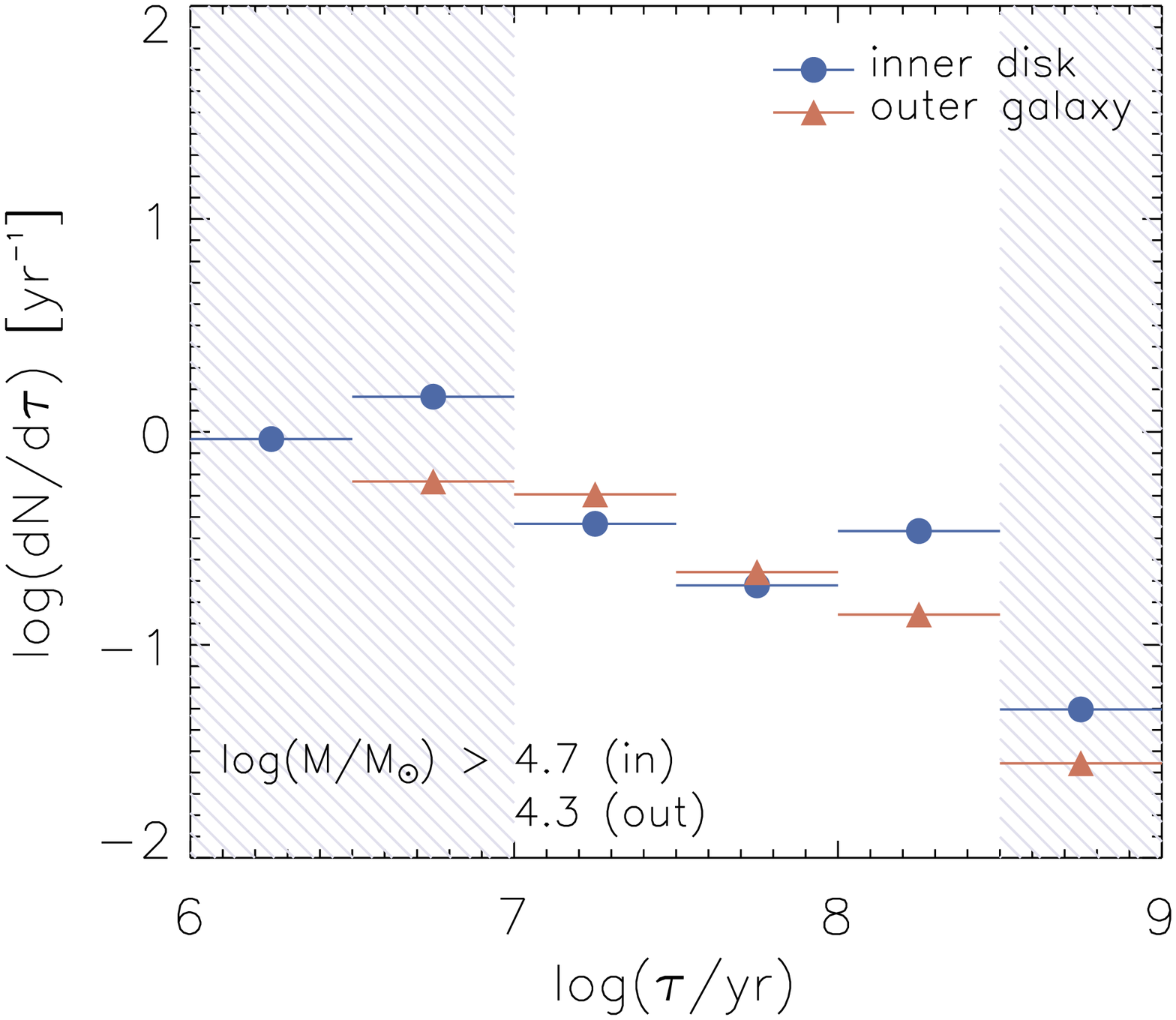}
		\includegraphics[width=0.49\textwidth]{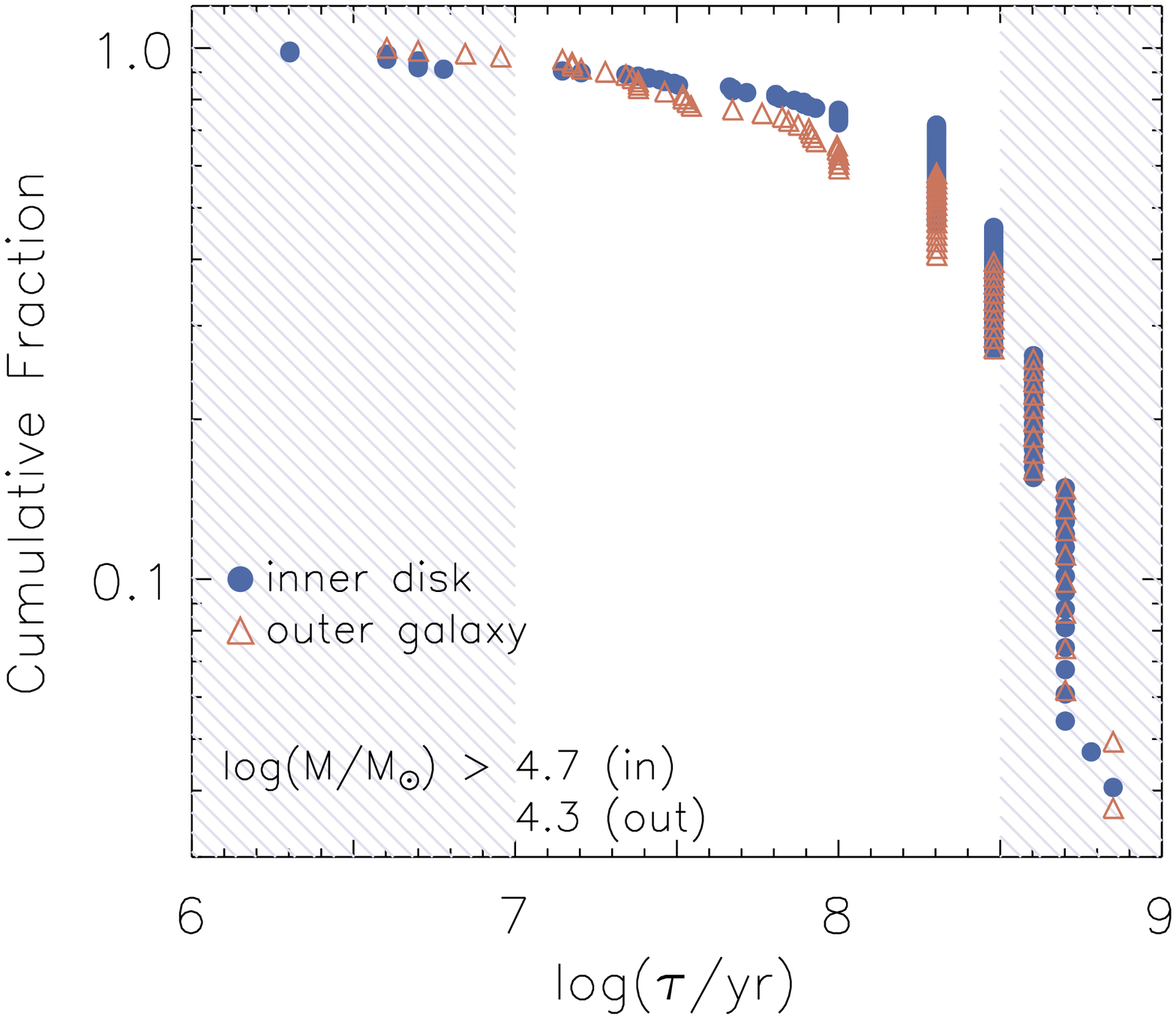}
		\caption{The cluster age distribution histograms 
		(`\dndt' plots) of the {\bf left panels} count the
		number of surviving clusters per time interval over 
		the past Gyr. Isotropic axes show an $x=-y$ slope in 
		the magnitude-limited sample on the top, as expected 
		from the literature. Applying a mass cut should flatten 
		the \dndt\ of a galaxy with a stable star formation 
		history. In this case, the slope lessens, but is 
		difficult to interpret given the small number of 
		data-points. 
		In this work we have argued that binning might erase
		information, therefore we also present unbinned data 
		in the cumulative age distribution on the ({\bf right}). 
		The two sets of plots show consistent results.  
		}\label{fig:dndt}

	%\end{center}
\end{figure*}

A final test for a physical difference between the inner/outer galaxy populations can be drawn from the cluster age distribution of Figure~\ref{fig:dndt} (left panels, often referred to as a `\dndt\ plot'), which counts the number of clusters surviving from every age-step of cluster formation \citep[see][and references therein]{lamers05, fall05, whitmore07, chandar10, bastian11}. When plotting all sources, as we do on the top panel, this diagnostic is affected by evolutionary fading and displays a power-law decline over time. Given a stable star formation history, setting a lower limit to the mass of a cluster sample will flatten the diagram to a certain age -- the higher the mass cut, the farther back in time this diagnostic can reach. 

In the bottom panel of Figure~\ref{fig:dndt} we perform this exercise for a mass cut that ensures completeness to $\log(\tau/\textup{yr})=8.5$ ($\approx300~$Myr). The precise mass of each cut (different for the inner and outer galaxy) is extracted from the age-mass diagram of Figure~\ref{fig:tM}, as the intersection of the $\log(\tau/\textup{yr})=8.5$ line and the detection limit. The hashed regions are meant to guide the eye away from certain regions: the first two bins contain not only clusters, but also unbound associations; and the last bin is affected by incompleteness. That precaution leaves only three bins from which to make a deduction, hence we refrain from quantitatively characterizing the star formation history of \n4041 from this \dndt\ plot. We note, however, that future SHUCS analyses focussed on nearby galaxies will not lack this diagnostic power.  

In previous Sections we have argued that binning the typically small data-sets of extra-galactic star cluster populations erases information contained in the individual data-points. In order to investigate this effect in the age distribution, we add two more plots to the right panels of Figure~\ref{fig:dndt}. There we modulate the age distribution by the mass of each cluster, to obtain a mass output plot. The full populations show a slight dissimilarity in their outputs over time, which, however, disappears when only plotting the complete sample of high-mass clusters. %This is consistent with the \dndt\ plots in showing a possible, gently increasing SFR over the past few hundred Myr in \n4041. 

% Figure: dN/dt
% ------------------------
%\vspace{-15pt}
% ------------------------

%%% SUBSECTION ----
% -----------------------------------------
\subsection{A Possible Dynamical Event, Revisited}\label{sec:sc:dyn}
In all, the study of star clusters in \n4041 is in tune with the derivations of the traditional, broad-band diagnostics of Section~\ref{sec:n4041}. The color difference between inner and outer galaxy is mirrored in the differing color, luminosity, mass, and age distributions of the corresponding star cluster populations. Unfortunately, the \dndt\ plot is limited by the number of clusters available above the 90\% completeness limit. Therefore, the past dynamical or accretion event  deduced through multi-wavelength metrics above cannot be confirmed through the cluster population. In following SHUCS investigations, however, we expect the smaller distances to the galaxies targeted to offer an opportunity to explore their dynamical histories via the age distribution.

%======= 7. SUMMARY =======
\section{Summary}\label{sec:summary}
We have presented the observational and technical setup of the Snapshot \textit{Hubble} U-band Cluster Survey (SHUCS), comprising new F336W (U-band equivalent) observations of 10 galaxies with existing \bvi\ coverage. We will use this new imaging to complete the \ubvi\ baseline, as required for the accurate photometric age dating of star clusters. The survey is focussed primarily on star clusters in the first $\sim$Gyr of their evolution, and is aimed at understanding the formation and early evolution of these objects, as well as their utility as tracers of star formation history. In this paper we have also demonstrated the high scientific yield of combining star cluster statistics with broad-band, multi-wavelength metrics. We propose that, with large-scale information readily available from \gal, SDSS, 2MASS, IRAS, and other surveys, this two-pronged approach is not only advantageous, but imperative in achieving a well-rounded understanding of each cluster population at hand. 

This combined approach was demonstrated through the analysis of the large-scale properties and stellar populations of \n4041, a massive galaxy at the upper tier of the SHUCS distance scale ($d\gtrsim15~$Mpc). This system features a complex physical and dynamical structure, expressed as a discontinuous brightness profile perhaps reminiscent of M64 from a different vantage point. We find a strong distinction in the colors and masses of star clusters when applying a cut in galactocentric distance where this break occurs, reminiscent of the color and age distributions in interacting and merging systems. Most notably, we find the inner disk to have been forming more massive clusters ($M\gtrsim10^4~$\Msun) in the past $\approx100~$Myr, despite the overall redder appearance of its subpopulation, which indicates an older mean age. This is in tune with our analysis of archival \gal\ FUV imaging, which reveals a higher SFR in the inner disk than the outer galaxy, similar to post-starburst systems. This could be linked to the theoretical expectation of a higher star formation efficiency and a higher fraction of stars forming in clusters in regions of higher gas density \citep{kruijssen12a}. We also discovered a tidal feature, which, combined with information in the literature, strongly favors an accretion event in the recent past ($\sim100~$Myr) as the origin of the segmented morphology of \n4041.

Throughout the galaxy we found a truncation in the star cluster mass function, in accord with the recent studies \citep{larsen09,gieles09a}. This truncation occurs at a higher mass in the inner disk, as expected by these recent results \citep{bastian12}. Furthermore, we found both the mass and luminosity function to break down into three segments when binning is abandoned in favor of cumulative distributions: one dominated by incompleteness, one following the familiar  power-law shape, and one encompassing a truncation at the upper end. Our results strongly advocate for the use of all available information, rather than binning, when characterizing the age, mass, and luminosity distributions of star cluster populations. In this work we have sought that result through the use of cumulative distributions. 

The strength of the survey-at-large derives from the availability of deep \uub\ observations at the highest resolution available (currently \hst-WFC3), enabling the precision age dating of hundreds of clusters per galaxy -- potentially thousands in nearby galaxies. The full survey will consist of the analysis of the cluster populations of ten late-type galaxies of various morphological and spectroscopic types. In addition to the unprecedented statistical value of this analysis, we expect to focus a few works on individual systems, namely the ongoing merger in \n2146 \citep[see the pilot study of][]{adamo12}, grand design spiral \n2997, and \n247, where we will contrast the star formation history derived from field stars to that derived from star clusters. Combined, the individual studies will help: 
\begin{enumerate}[itemsep=0pt]
	\item Search for a characteristic value ($M_*$) in the cluster 
			mass function. 
	\item Empirically constrain star cluster disruption laws. 
	\item Constrain the fraction of stars that form in clusters.
	\item Plot the star/cluster formation histories of SHUCS 
			galaxies over the past Gyr. 
	\item Determine whether cluster size is affected by local 
			conditions. 
	\item Search for environmental dependencies in all of the above. 
\end{enumerate}

\acknowledgements We are grateful to the anonymous referee for the constructive commentary. Support for this work was provided by NASA through grant number HST-GO-12229.01-A from the Space Telescope Science Institute, which is operated by AURA, Inc., under NASA contract NAS5-26555. SSL acknowledges the hospitality of the International Space Science Institute (ISSI) in Bern, Switzerland. EZ acknowledges research grants from the Swedish Research Council and the Swedish National Space Board. The research leading to these results has received funding from the European Community's Seventh Framework Programme (/FP7/2007-2013/) under grant agreement No 229517. 
This work is partly based on observations made with the NASA Galaxy Evolution Explorer. GALEX is operated for NASA by the California Institute of Technology under NASA contract NAS5-98034.
This paper makes use of publicly available SDSS imaging and spectroscopy. Funding for the creation and distribution of the SDSS Archive has been provided by the Alfred P. Sloan Foundation, the Participating Institutions, the National Aeronautics and Space Administration, the National Science Foundation, the U.S. Department of Energy, the Japanese Monbukagakusho, and the Max Planck Society. The SDSS Web site is http://www.sdss.org/. The SDSS is managed by the Astrophysical Research Consortium (ARC) for the Participating Institutions. The Participating Institutions are The University of Chicago, Fermilab, the Institute for Advanced Study, the Japan Participation Group, The Johns Hopkins University, Los Alamos National Laboratory, the Max-Planck-Institute for Astronomy (MPIA), the Max-Planck-Institute for Astrophysics (MPA), New Mexico State University, Princeton University, the United States Naval Observatory, and the University of Washington.
This research has made use of the NASA/IPAC Extragalactic Database (NED) which is operated by the Jet Propulsion Laboratory, California Institute of Technology, under contract with the National Aeronautics and Space Administration. 

%======= BIBLIOGRAPHY =======
\bibliographystyle{apj}
\bibliography{./references}

\end{document}

%% file: tab_LF.tex
\begin{table*}[!p]
    \caption{Luminosity Function Fit Parameters and Outcomes.\label{tab:LF}}
    \begin{center}
    \begin{tabular}{llccccc}
    \hline
    \hline
    Filter & \tmult{Fit~Range~(in/out)} & \tmult{Slope, unbinned~(in/out)} & \tmult{Slope, binned~(in/out)} \\
           &     \tmult{(mag)}          & \tmult{$(-)$}                    & \tmult{$(-)$} \\
%    \hline % flat, in/out consistent
%    F336W  & flat segment & $[20.0, 22.5]$&$[20.0, 22.5]$ & $2.27\pm0.18$&$2.48\pm0.39$ & \\
%    F450W  & flat segment & $[21.0, 23.0]$&$[21.0, 23.0]$ & $2.69\pm0.37$&$2.63\pm1.09$ & \\
%    F606W  & flat segment & $[21.0, 23.0]$&$[21.0, 23.0]$ & $2.52\pm0.39$&$3.25\pm0.96$ & \\
%    F814W  & flat segment & $[21.0, 23.0]$&$[21.0, 23.0]$ & $2.25\pm0.33$&$3.19\pm0.69$ & \\
    \hline % flat segment, UBVI consistent
    F336W~(U) & $[20.5, 22.0]$&$[21.5, 23.5]$ & $2.31\pm0.24$&$2.56\pm0.15$ & $2.35\pm0.10$&$2.56\pm0.10$\\
    F450W~(B) & $[20.5, 22.0]$&$[21.5, 23.5]$ & $2.94\pm0.57$&$2.84\pm0.56$ & $3.21\pm0.12$&$2.94\pm0.22$\\
    F606W~(V) & $[20.5, 22.0]$&$[21.5, 23.5]$ & $3.13\pm0.40$&$3.06\pm0.48$ & $3.10\pm0.13$&$3.37\pm0.14$\\
    F814W~(I) & $[20.5, 22.0]$&$[21.5, 23.5]$ & $3.07\pm0.37$&$2.93\pm0.21$ & $2.29\pm0.11$&$2.68\pm0.05$\\
    \hline % full range
    F336W~(U) & $[20.0, 23.6]$&$[20.0, 24.6]$ & $2.01\pm0.21$&$2.23\pm0.23$ & $1.60\pm0.04$&$1.84\pm0.06$\\
    F450W~(B) & $[20.0, 23.6]$&$[21.0, 24.6]$ & $2.46\pm0.38$&$2.81\pm0.16$ & $2.23\pm0.06$&$2.55\pm0.04$\\
    F606W~(V) & $[20.0, 23.6]$&$[21.0, 24.6]$ & $2.37\pm0.42$&$2.76\pm0.25$ & $1.90\pm0.07$&$2.60\pm0.06$\\
    F814W~(I) & $[20.0, 23.6]$&$[21.0, 24.6]$ & $2.27\pm0.43$&$2.57\pm0.27$ & $1.56\pm0.07$&$2.17\pm0.06$\\
    \hline
    \end{tabular}
    \end{center}
    \label{tab:LF}
	\tablecomments{The top and bottom tiers show fits to the `smooth' and `full' ranges (see text). The limits 
		of the full range fits measure between 20 mag and the 90\% completeness limits for the inner disk and 
		outer galaxy (see Section~\ref{sec:completeness}).}
\end{table*}%

% LF slopes and fit ranges: 
% 
% 1) Full range: 
%
%   lim336 = [20.0, comp90in, 20.0, comp90out] mag
%   lim450 = [20.0, comp90in, 21.0, comp90out] mag
%   lim606 = [20.0, comp90in, 21.0, comp90out] mag
%   lim814 = [20.0, comp90in, 21.0, comp90out] mag
% 
% (comp90in  = 23.6 mag)
% (comp90out = 24.6 mag)
% 
%            --F336W:  2.01 +/- 0.21,  2.25  +/- 0.23
%            --F450W:  2.46 +/- 0.38,  2.81  +/- 0.16
%            --F606W:  2.37 +/- 0.42,  2.76  +/- 0.25
%            --F814W:  2.27 +/- 0.43,  2.57  +/- 0.27
% 
% 2) Flat part only
% 
%   lim336 = [20.0, 22.5, 20.0, 22.5] mag
%   lim450 = [21.0, 23.0, 21.0, 23.0] mag
%   lim606 = [21.0, 23.0, 21.0, 23.0] mag
%   lim814 = [21.0, 23.0, 21.0, 23.0] mag
%
%            --F336W:  2.27 +/- 0.18,  2.48  +/- 0.39
%            --F450W:  2.69 +/- 0.37,  2.63  +/- 1.09
%            --F606W:  2.52 +/- 0.39,  3.25  +/- 0.96
%            --F814W:  2.25 +/- 0.33,  3.19  +/- 0.69